# The impact of generative artificial intelligence on socioeconomic inequalities and policy making

Valerio Capraro[1,*], Austin Lentsch[2], Daron Acemoglu[3], Selin Akgun[4], Aisel Akhmedova[4], Ennio Bilancini[5], Jean-François Bonnefon[6], Pablo Brañas-Garza[7], Luigi Butera[8], Karen M. Douglas[9], Jim A.C. Everett[10], Gerd Gigerenzer[11], Christine Greenhow[12], Daniel A. Hashimoto[13,14], Julianne Holt-Lunstad[15], Jolanda Jetten[16], Simon Johnson[17], Werner H. Kunz[18], Chiara Longoni[19], Pete Lunn[20], Simone Natale[21], Stefanie Paluch[22], Iyad Rahwan[23], Neil Selwyn[24], Vivek Singh[13], Siddharth Suri[25], Jennifer Sutcliffe[4], Joe Tomlinson[26], Sander van der Linden[27], Paul A. M. Van Lange[28], Friederike Wall[29], Jay J. Van Bavel[30, 31], Riccardo Viale[32]

[1] Department of Psychology, University of Milan-Bicocca, Italy. [2] Department of Economics, MIT, USA. [3] Institute Professor and Department of Economics, MIT, USA. [4] College of Education, Michigan State University. [5] IMT School of Advanced Studies Lucca, Italy. [6] Toulouse School of Economics, France. [7] Department of Economics, Loyola Andalucia University, Spain. [8] Department of Economics, Copenhagen Business School, Denmark. [9] School of Psychology, University of Kent, UK. [10] School of Psychology, University of Kent, UK. [11] Max Planck Institute for Human Development, Berlin, Germany. [12] College of Education, Michigan State University. [13] Penn Computer Assisted Surgery and Outcomes Laboratory, Department of Surgery, Perelman School of Medicine, University of Pennsylvania. [14] Department of Computer and Information Science, School of Engineering and Applied Science, University of Pennsylvania. [15] Department of Psychology and Neuroscience, Brigham Young University, USA. [16] School of Psychology, University of Queensland, Australia. [17] MIT Sloan School of Management, USA. [18] Department of Marketing, University of Massachusetts Boston, USA. [19] Department of Marketing, Bocconi University, Italy. [20] Economic & Social Research Institute, Dublin, Ireland. [21] Department of Humanities, University of Turin, Italy. [22] Department of Service and Technology Marketing, Aarhus University, Denmark. [23] Center for Humans and Machines, Max Planck Institute for Human Development, Berlin, Germany. [24] Faculty of Education, Monash University, Australia. [25] Microsoft Research, USA. [26] York Law School, United Kingdom. [27] Department of Psychology, University of Cambridge, UK. [28] Department of Experimental and Applied Psychology, Vrije Universiteit, Amsterdam, The Netherlands. [29] Department of Management Control and Strategic Management, University of Klagenfurt, Austria. [30] Department of Psychology & Center for Neural Science, New York University, USA. [31] Norwegian School of Economics, Bergen, Norway. [32] Department of Economics, University of Milan-Bicocca, Italy.

* Corresponding author: valerio.capraro@unimib.it

**Abstract**

Generative artificial intelligence has the potential to both exacerbate and ameliorate existing socioeconomic inequalities. In this article, we provide a state-of-the-art interdisciplinary overview of the potential impacts of generative AI on (mis)information and three information-intensive domains: work, education, and healthcare. Our goal is to highlight how generative AI could worsen existing inequalities while illuminating how AI may help mitigate pervasive social problems. In the *information* domain, generative AI can democratize content creation and access, but may dramatically expand the production and proliferation of misinformation. In the *workplace*, it can boost productivity and create new jobs, but the benefits will likely be distributed unevenly. In *education*, it offers personalized learning, but may widen the digital divide. In *healthcare*, it might improve diagnostics and accessibility, but could deepen pre-existing inequalities. In each section we cover a specific topic, evaluate existing research, identify critical gaps, and recommend research directions, including explicit trade-offs that complicate the derivation of a priori hypotheses. We conclude with a section highlighting the role of policymaking to maximize generative AI's potential to reduce inequalities while mitigating its harmful effects. We discuss strengths and weaknesses of existing policy frameworks in the European Union, the United States, and the United Kingdom, observing that each fails to fully confront the socioeconomic challenges we have identified. We propose several concrete policies that could promote shared prosperity through the advancement of generative AI. This article emphasizes the need for interdisciplinary collaborations to understand and address the complex challenges of generative AI.

# Introduction

*"The rise of powerful AI will be either the best, or the worst thing, ever to happen to humanity. We do not yet know which."* - Stephen Hawking, 2016

Advances in generative Artificial Intelligence (AI) represent a shift in the capability of these systems to solve problems previously thought unsolvable (Bucker et al., 2023). Techno-optimists predict a utopian future where machines can perform an ever-increasing number of tasks—but humans remain in control, the gains from prosperity are shared throughout society, and we all enjoy lives with less work and more leisure. In contrast, pessimists forecast a dystopian future where machines not only replace humans in the workplace, but also surpass human capability and oversight, destabilize institutions and destroy livelihoods—and perhaps even cause the downfall of humanity (Bostrom, 2017).

Melvin Kranzberg, a prominent scholar in the history of technology, in a presidential address to his field, defined "Kranzberg's Laws", the first of which states that *"Technology is neither good nor bad; nor is it neutral"* (Kranzberg, 1985). This principle suggests that technologies like generative AI will likely have negative and positive impacts on society, though they are not inherently predestined toward either. In the current article, we outline some of the more likely positive and negative effects, with the aim of guiding scholars and policy makers to harness the positive potential of this new technology while mitigating the costs for individuals and society.

Both optimists and pessimists agree that generative AI represents a qualitative departure from previous automation processes, such as microelectronics, information technology, and the Internet. Unlike traditional automation, which primarily focuses on replicating predefined tasks, generative AI introduces the ability to create new, original output. The implications have the potential to reshape foundational values and skills. For instance, while generative AI might facilitate written communication, especially for non-native speakers, it could devalue foundational language learning. The incentives to master syntax, vocabulary and grammar might wane as generative AI begins to exceed the skill level of humans. This shift reflects a broader theme: generative AI does not merely alter practices but fundamentally transforms the valuation of knowledge and skills (Gigerenzer, 2022).

The effects of generative AI may eventually expand to virtually every facet of society (Dwivedi et al., 2023). We begin by discussing the impact of generative AI in the information domain. Generative AI can democratize content creation and access to information, but could also lead to challenges of increased misinformation and eroded trust in digital content.

In the subsequent sections, we investigate potential downstream impacts on socioeconomic inequalities in three key information-intensive areas: work, education, and healthcare. In the *workplace*, generative AI could increase productivity and promote shared prosperity, especially when used to complement human efforts and create new well-paid jobs. However, the benefits and costs will likely be distributed unevenly across firm sizes, sectors, and worker demographics. In *education*, generative AI promises personalized learning experiences, potentially bridging educational gaps. However, it also raises concerns about equal access to these advanced tools. The *health* sector could greatly benefit from AI's diagnostic and predictive capabilities, improving patient outcomes and making healthcare more accessible. Yet, there is the risk of deepening existing inequalities of care and access, especially for under-resourced and marginalized communities. For each of these domains, we explore current research and suggest future directions.

We conclude with an examination of the role of policymaking in the age of AI. We discuss the pros and cons of the current policy approaches in the European Union, the United States, and the United Kingdom, noting that all fall short in addressing the socioeconomic risks that we identify. We argue that policies must be designed to mitigate the potential problems posed by AI, without increasing inequality and harm to vulnerable members of society. We recommend several policies that should be studied empirically and included in public debate. These include measures to combat AI-generated misinformation, prevent job market inequalities, and bridge the digital divide in education and healthcare. The goal should be to harness the potential of generative AI in ways that favor human flourishing, striking a balance between technological advancement and societal well-being.

## Impact on (mis)information

Generative AI has the potential to revolutionize the information domain, impacting areas such as work, education, healthcare, law, finance, and policy making. One advantage is the ability for personalization, where AI can tailor content to individual preferences, enhancing and customizing user experiences. The language translation and localization capabilities of AI extend the reach of content globally, crossing traditional language barriers and making material accessible to a wide variety of cultural contexts and social groups (Rathje et al., 2023a). For example, AI aids in making information more accessible for individuals with disabilities, by creating text-alternative formats like audio or simplified summaries. AI may also help automate the fact-checking process, aiding the spread of accurate information (Hoes et al., 2023; Zavolokina et al. 2024).

Concerningly, new generative AI technology and sophisticated machine learning techniques may also enable companies to collect and deploy excessive amounts of information about individuals. This will enable exploitation of consumers' biases or vulnerabilities in order to capture more of the consumer surplus via price discrimination or violations of consumer privacy, leading to "surveillance capitalism" (Acemoglu, 2024; Zuboff, 2023). A dominant model has emerged from these monopolies, where internet platforms earn income by optimally marketing digital advertisements (Acemoglu and Johnson, 2023). This strategy places a premium on user attention, which has led companies to deploy AI and machine learning techniques to prolong user engagement, often to the detriment of individual and societal well-being (Wu, 2016; Acemoglu, Ozdaglar, et al., 2023; Acemoglu, 2023; Rathje et al., 2023b). Relatedly, companies with more data possess an anticompetitive advantage, enabling them to exercise market power to extract surplus and relax price competition, which can be detrimental for consumers (Acemoglu, 2024).

Malicious actors can exploit generative AI to create false information in ways that convincingly copy the style and content of human-created text, by synthetically generating text, audio, images, and videos ("deepfakes"). For instance, malicious generative AI tools like WormGPT (a ChatGPT alternative for designing and refining cyber-attack strategies and malware) or PoisonGPT (a modified open-source AI model designed to spread misinformation) show that these tools can be used to accomplish malign aims and to sabotage further technology development.

Manipulated political images already make up a substantial portion (~20%) of visual misinformation on social media (Yang et al., 2023). This type of misinformation can be especially common during elections and intergroup conflicts such as the Russo-Ukrainian and Israel-Gaza wars (Tworney et al., 2023). Additionally, people are largely unable to tell the difference between AI- and human-generated text (Kreps et al., 2022) and that AI has been shown to generate more convincing misinformation than humans (Spitale et al., 2023) as well as persuasive propaganda (Goldstein et al., 2024). Consequently, there is growing concern that generative AI may also increase the quantity of misinformation. Indeed, hundreds of unreliable AI-news websites have popped up (Newsguard, 2023). There is currently little legislation preventing the use of deepfakes in political campaigns, although there are some steps in this direction. Some US states have introduced legislation prohibiting their use (Beller, 2024). The EU AI Act (Article 52(3)) does not outlaw deepfakes, but at least requires platforms to identify AI-generated content.

AI-assisted misinformation can spread rapidly on social media and micro-targeting people with deepfakes can influence their attitudes toward politicians (e.g., Dobber et al., 2021). GPT can also automate micro-targeting in a way that makes it more persuasive than standard non-personalized ads (Simchon et al., 2024). The possibility to create information that is personalized or targeted to specific individuals and groups is likely to increase, especially during elections (Benson, 2023). Politicians, including Republican presidential candidate Ron DeSantis, have already started using deepfakes in their political campaigns, such as fake images of Donald Trump hugging Anthony Fauci (McCarthy, 2023). The 2024 US Presidential election has been dubbed "The Deepfake Election" (Wall Street Journal, 2024).

This increase in misinformation may have significant social consequences. Conspiracy theories and misinformation can contribute to attitude polarization (Del Vicario et al., 2016) and undermine trust (Tworney et al., 2023). Moreover, political conspiracy theories and misinformation can affect voting decisions, health-related conspiracy theories can influence people's medical choices (e.g., vaccination), and misinformation and conspiracy theories can fuel conflict between groups (Douglas, 2021; Sternisko et al., 2020). While some people simply ignore online misinformation (Acerbi et al., 2022), this content is likely to penetrate specific groups, especially since AI may help automate the micro-targeting process in which thousands of persuasive messages can now be generated easily at scale (Simchon et al., 2024). For example, there is evidence that Trump voters were more susceptible to misinformation during the 2016 presidential election (Grinberg et al., 2019).

Therefore, regulation and interventions are urgently needed to limit the diffusion of AI-generated misinformation. Simply warning people of deepfakes or including a tag clarifying whether a piece of content is AI-generated might backfire, as such tags may reduce the believability of true content as well (Longoni et al., 2022; Tworney et al., 2023). In the realm of human-generated misinformation on social media, psychological interventions based on accuracy-salience and educational interventions based on inoculation theory improve the quality of information shared. For example, making the concept of accuracy salient can reduce the sharing of fake news, without adversely affecting the dissemination of accurate news (Pennycook and Rand, 2022). Moreover, endorsing accuracy not only decreases the sharing of false news, but also increases the sharing of true news (Capraro and Celadin, 2023). However, these interventions have typically modest effects compared to alternative strategies (Pretus et al., 2024).

Inoculation theory or "prebunking" is a preemptive approach to countering misinformation that follows the vaccination analogy (van der Linden, 2023). Several inoculation games and videos have been developed to expose subjects to controlled (weakened) doses of misinformation along with tools on how to spot it and these activities make them better at detecting online manipulation (Roozenbeek and Van der Linden, 2019). Similarly, in a field study on YouTube, videos containing micro-doses of common misinformation techniques increased discernment of online manipulation tactics (Roozenbeek et al., 2022). Because prebunking often works better than debunking (Jolley and Douglas, 2017), future work could adapt these techniques to AI-generated news (Shin and Akhtar, 2024).

One concern, however, is that interventions based on accuracy-salience and inoculation may be most effective for easily discernible misinformation. Unfortunately, generative AI makes misinformation more subtle and harder to discern (Kreps et al., 2022), which may necessitate a new toolbox of interventions, specifically designed to counteract (visual) AI-generated misinformation. Moreover, generative AI could lead to entirely new challenges, such as tackling misinformation disseminated via one-to-one personalized communications (e.g., through bots). This further highlights the urgency to adapt existing or develop a new set of intervention strategies (Feuerriegel et al., 2023). New policies on social media platforms and effective regulation are likely needed to address this issue at scale rather than relying too heavily on subtle interventions. An intriguing direction would be to explore how generative AI itself could be leveraged to combat misinformation. For instance, one study found that engaging in dialogue with AI reduces conspiracy beliefs among conspiracy believers (Costello et al., 2024).

Even in the absence of malicious actors, the most advanced AI-systems are known to "hallucinate" false information in a very realistic manner (Bubeck et al., 2023). These hallucinations may induce complex social dynamics, like self-fulfilling prophecies, where an initially false prediction becomes true just because someone—e.g., a generative AI system— asserts that it will become true (Merton, 1948). In this sense, AI may produce prophecies that could "take a life for their own" (Citron and Pasquale, 2014). For example, automated scoring systems that predict the likelihood of default on debt repayment may contribute to (or even cause) credit risk.

Another important area of concern is that individuals may not know whether they are interacting with a person or a machine (Natale, 2021). This is increasingly likely to be the case when people engage online with businesses and public services. If they believe, rightly or wrongly, that they are interacting with a machine, their behavior is likely to change (March, 2021). For instance, while AI can generate responses that make people feel heard by offering emotional support, people feel more heard when they believe a response comes from a human (Yin et al., 2024). The mere knowledge people are interacting with a machine can change their experience.

Human behavior tends to become more selfish in human-machine interactions because reciprocation—a vital factor in sustaining prosocial behavior—is not maintained as consistently as in human-human interactions (Ishowo-Oloko et al., 2019; Makovi et al., 2023). Prosocial behavior depends on people's beliefs about the relationship between the machine and the humans behind it (von Schenk et al., 2023). When interacting with a machine, people respond less emotionally, feeling less guilt about being ungenerous (Chugunova and Sele, 2022). They become more likely to be dishonest in pursuit of monetary rewards (Cohn et al., 2022). An outstanding question concerns whether similar slippage from ethical standards occurs not only among people interacting with a machine, but also among those who delegate to the machine (Köbis et al., 2021).

One overlooked implication is the impact of generative AI on the plurality of information available on the web. Companies including Microsoft and Google have envisioned integrating large language models into their search engines, but the implications of this move have only started to be explored (Cutler, 2023). Among the most significant implications is users' access to information. Search engines powered by generative AI may restrict the plurality of information available on the Web. When users input a query into the current version of Google Search they are pointed to a plurality of sources. Although users tend to select among the first results returned by the search engine (Goldman, 2008), the interface enables them to browse many alternative results. The same input directed to a search engine powered by generative AI will provide an extra layer of mediation likely to provide a much more limited amount of source information, unless specific design features are included to counteract this (Natale and Cooke, 2021).

In addition to reducing access to information, generative AI may also threaten the quality and availability of online information. The already-pervasive issue of bot accounts may be exacerbated by new generative technologies, which can assist in coding a multiplicity of these bots as well as providing text content for the bots to post (Ferrara et al., 2016). These tools could also be used to generate content optimized for search engines *en masse*, a useful tactic for businesses to "poach" traffic from competitors' websites (e.g., Semrush, 2023).

This is a problem because current generative AI tools essentially provide an "average" response to a particular question, and these models are trained largely on text data collected from the internet; therefore, if the practice of generating content optimized for search engines at massive scale becomes common practice, both generative AI tools and online information may crater into an average of averages, lacking true insight, creativity, or novel ideas. At the very least, it could make useful contributions difficult to identify within a sea of mediocre machine-generated "average" content.

An analogous concern is that common message boards and websites for knowledge sharing (e.g., Stack Overflow) have experienced both a reduction in questions posted—especially the basic questions that ChatGPT does well at answering—and an increase in question responses, perhaps due to writing aid from tools like ChatGPT (del Rio-Chanona et al., 2023). Though these Q&A sites require competent subject-matter experts to provide insights and suggestions, they also require neophytes to ask those questions in the first place. Reduced engagement by novice users not only has effects on the continued usefulness of these websites to aggregate and share knowledge, but also for innovation and creativity that may rely on content from these platforms as input.

Many of these challenges require governmental regulation. We will discuss specific policy recommendations in the last section. However, it is important to recognize that organizations also play a significant role. Organizations often face trade-offs between achieving their profitability goals and adhering to ethical practices. One way to view these trade-offs is through a utilitarian perspective, weighing the benefits and costs. For instance, privacy experts have developed the "privacy calculus" to address such trade-offs from the consumer perspective (Laufer & Wolfe, 1977). Similarly, Wirtz et al. (2023) argue that service firms engage in a "corporate digital responsibility calculus" and weigh the benefits and costs of following ethical principles to determine their level of engagement in digital responsibility. Therefore, we might expect service firms to engage in responsible practices when the benefits outweigh the costs. Digital responsibility does present benefits. By adopting responsible practices, corporations can build trust with their customers and stakeholders, differentiate themselves from competitors, and enhance their reputation. However, if the costs of adopting responsible practices is too high, firms may be less likely to engage in good practices and regulatory enforcement may be necessary (Wirtz et al., 2023).

In conclusion, while generative AI has the potential to expand access to and content of information, it also raises significant challenges such as market anticompetitive advantages, data misuse, data poisoning, misinformation proliferation, and altered human-machine interactions, all of which necessitate careful consideration and targeted research. Table 1 summarizes the main directions of future research, along with one specific research question for each direction, potential design, and theoretical trade-off that complicate the derivation of hypotheses a priori. Supplementary Table S1 extends this table to three specific research questions for each direction. This list is not meant to be exhaustive but serves as an initial guide for subsequent investigations.

| | | | |
|---|---|---|---|
| **Future research directions** | | | |
| **Research area** | **Specific question** | **Potential design** | **Trade-off** |
| Investigate how AI can be used to make information more accessible, especially for individuals with disabilities. | Can AI-based summarization tools improve information accessibility for individuals with cognitive disabilities by simplifying complex texts? | Evaluate the comprehension of complex news articles by individuals with cognitive disabilities after using AI-based summarization tools, compared with individuals not using these tools. | Summarization could make content more accessible, but oversimplification might omit critical details, challenging the balance between accessibility and content accuracy. |
| Understand how the largest firms could monopolize the future of AI; find ways for smaller and innovative firms to effectively compete with those largest players. | Does open-source AI decrease the risk of monopolization by large firms? | Compare the growth and success rates of firms that contribute to or use open-source AI versus those relying on proprietary solutions. | Open-source could democratize AI development, but the capacity of small firms to benefit from it may be smaller compared to large firms', affecting the potential of such measures to reduce the competitive gap. |
| Explore regulatory measures to prevent misuse or inappropriate access to data by AI systems. | Can a standardized transparency protocol prevent data misuse and privacy breaches? | Develop a standardized transparency protocol and test its impact on the frequency of data misuse and privacy breaches. Measure also users' trust in the system and data sharing. | Increased transparency may decrease data misuse and privacy breaches, but the difficulty of transparency protocols may actually decrease public trust and data sharing. |
| Investigate strategies to identify and limit the spread of | Can dialoguing with gen AI reduce beliefs in misinformation? | Compare misinformation beliefs between participants engaging with an AI- | AI's capabilities for personalization, linguistic fluency, and logical reasoning can |
| misinformation generated by AI. | | tool designed to counter misinformation and those interacting with a neutral AI. | diminish misinformation beliefs, but the risk of hallucinations could generate new misinformation. |
| Explore ways to design AI-systems that support cooperative and ethical behavior in human-machine interactions. | Do humans become more unethical when they can delegate decisions to AI compared to other humans? | Subjects make morally ambiguous decisions, with the option to delegate these decisions either to AI or to another human, observing changes in their ethical standards. | The emotional detachment of AI may reduce moral responsibility, but the potential for AI to document and expose decisions might increase personal accountability. |
| Examine how AI-enhanced search engines can be designed to preserve user autonomy and plurality of information. | Does making the data sources of AI-enhanced search engines transparent influence user engagement with alternative information sources? | Disclose the data sources for AI algorithms in search engines and assess whether users seek out additional, alternative sources of information as a result. | Transparency about data sources might encourage users to seek diverse information, but could also lead to information overload. |
| Consider how the proliferation of AI-generated content could lower the quality of online information and ensure that human users can continue to contribute new knowledge. | How does the ratio of AI-generated to human-generated content affect content diversity on online forums? | Manipulate the ratio of AI-generated to human-generated content in an online forum environment. Survey users on perceived content diversity and quality. | Higher ratios of AI content could enhance content availability, but may also homogenize the content or lead to mediocre or over-creation of banal, "average" content. |
| Investigate the role of Corporate Digital Responsibility and its implementation challenges | What are the main barriers to implementing effective digital responsibility strategies in organizations? | Development of organizational strategies to implement a general Corporate Digital Responsibility framework. | Many organizations might not be aware of the role and importance of corporate digital responsibility strategies, which |

| | | | might cause over-regulation. |
|---|---|---|---|
| | | | |

*Table 1. Summary of the main research directions on the impact of generative AI on information. For each relatively broad research area, we propose three specific research questions, along with potential experimental designs. We also identify an underlying theoretical trade-off that complicates the derivation of a priori hypotheses. This list is not exhaustive but rather provides examples of the kinds of questions future research should aim to address.*

# Impact on work

Previous waves of digital technologies have contributed to increased inequality. Some of these technologies, like personal computers, have been complementary mostly to educated workers (Autor et al., 1998; Autor, 2019), while others, like industrial robots, have been used to automate repetitive or systematic tasks that are often performed by less-educated workers (Acemoglu and Restrepo, 2022a, 2022b). Together, the upside for more-educated workers and downside for less-educated workers have magnified the distributional consequences of technological innovation, highlighting what is sometimes called "Skill-Biased Technological Change" (Acemoglu, 2002).

The current trend in AI emphasizes automation. While some amount of this is unavoidable, the displacement of labor by "so-so technologies" (e.g., self-checkout kiosks or automated phone systems) that offer little or no productivity gain, along with diminished worker voice due to intensified monitoring and surveillance, can be harmful to long-run productivity and other social goals like job satisfaction (Acemoglu and Restrepo, 2022a, 2019). Although new technologies can boost productivity (Brynjolfsson and McAfee, 2016), the gains have often fallen below expectations, especially when the focus has been on replacing work instead of augmenting worker capabilities or developing new ones (Acemoglu et al., 2016; Acemoglu and Johnson, 2023).

New technologies like AI should be oriented not so much toward replacing human problem-solving abilities, but rather toward enhancing them in a symbiotic relationship where machines are designed to complement human capabilities and humans can compensate for the weaknesses of machines (Licklider, 1960). This "pro-worker" or "human-complementary" path could contribute more to productivity growth and could help reduce economic inequality. The question we is whether AI will accelerate the existing trend of automation without the offsetting force of good-job creation—particularly for non-college educated workers—or whether it will instead introduce new value-adding tasks and well-paying jobs for workers with diverse skill sets and educational backgrounds.

There is cause for optimism: AI can complement workers by making them more efficient, helping them to produce higher quality work, or enabling them to take on new value-adding tasks (Acemoglu and Restrepo, 2018; Alam et al., 2023; Korinek, 2023). Brynjolfsson and et al. (2023), for instance, consider the staggered implementation of a chat assistant by a Fortune 500 software company that provides business process software. The chat assistant monitored customer service chats and proposed real-time response suggestions to customer service agents. Less-skilled or inexperienced workers resolved around 34% more issues per hour, with average improvement across all workers measuring about 14%. Agents using the tool with only two months of tenure performed as well as those without the tool who had more than six months of tenure.

Another study examined the impact of GPT-4 access on complex knowledge-intensive tasks. AI users were more productive and produced higher quality work. However, for tasks beyond the capabilities of GPT-4—specifically, tasks that involve imperfect information or omitted data, which require cross-referencing resources and leveraging experience-gained intuition—AI usage resulted in fewer correct solutions. Consultants with below-average performance improved by 43% with AI, while those above average improved by 17% (Dell'Acqua et al., 2023).

Similar patterns have been observed in other studies. For instance, Peng and co-authors (2023) conducted a controlled experiment with GitHub Copilot, an AI-based programming assistant. Programmers with access to the AI copilot completed a task in 71 minutes on average, less than half the time of the control group's 161 minutes. The AI assistant provided the biggest boost to less-experienced and older programmers, as well as those coding more hours daily. Similarly, people with access to ChatGPT completed a writing task faster and produced higher quality work (Noy and Zhang, 2023). Again, this reduced worker inequality by benefiting lower-ability workers more; moreover, it led to higher job satisfaction and self-efficacy.

These studies underscore the potential of generative AI to disproportionately boost productivity for workers with less experience or skill. This differs from previous technologies. Instead of mostly benefiting more-skilled workers, generative AI tools seem to exhibit a worker-complementary "inverse skill-bias," benefitting *less-skilled workers* much more than highly skilled workers. An example discussed further in the healthcare section finds that some physicians perform more poorly when they use AI tools to support diagnostic decisions, compared to their performance without AI usage (Agarwal et al., 2023). This suggests that gains from integrating generative AI into medicine could be better targeted to nurses, medical technicians, or healthcare administrators (Acemoglu, Autor et al., 2023). These workers have the requisite baseline knowledge to conduct critical components of care delivery, but perhaps could be empowered to do alleviate the "enormous administrative burdens" that currently exacerbate physician shortage (Ehrenfeld, 2023). These AI tools and practical applications are still in their infancy. Yet, if these trends continue, it is possible that generative AI could reverse the income and job market inequalities and "rebuild the middle class" in advanced economies (Autor, 2024).

Generative AI could also reduce barriers to entry in the digital economy. For example, its translation capabilities can help overcome language barriers. This increased accessibility, in conjunction with trends toward diminishing geographic barriers, could have a compounding positive effect. Accordingly, there has been a surge of interest in remote-enabled digital economy jobs, especially in rural areas (Counts et al., 2022). Yet, these benefits have mostly favored well developed countries. While generative AI could also offer expanded opportunities to countries in the "Global South", it is unlikely to have much direct impact in the near term due to insufficient investment in prerequisite digital infrastructure, local researchers, and broader digital skills training (Mannuru et al., 2023).

One strength of generative AI is its ability to parse and aggregate enormous amounts of information. This capability can equalize access to information and lower research costs by simplifying online search tasks. If a user wants to accomplish a complex task with a traditional search engine, they have to break that task into pieces, issue search queries for each piece, read the web pages returned by the search engine, assess the representativeness of their gathered information, and then aggregate the results to solve the problem. Generative search engines, on the other hand, can aggregate this information and return it to the user, requiring less bandwidth and fewer trips between the user and the system which would be helpful in lower resource environments. In addition to the time and cost savings, these tools could compensate for expertise by identifying trustworthy resources and extracting the consensus on any topic by simultaneously considering more information than human operators can retain. This approach could help users and businesses in low-resource settings access information that has traditionally been available only in high-resource environments.

However, there are also ways in which AI might exacerbate inequalities in the workplace. One concern is differential access to these tools. The most widely available and accessible generative AI platforms still require additional technical inputs (e.g., internet access and internet-enabled devices) as well as training to optimize performance. Industries, firms, and workers that have not yet integrated the prerequisite technologies will struggle to take advantage of the expanded capabilities and consequent productivity and earnings upsides, likely falling (further) behind well-resourced competitors or coworkers.

The role of firm behavior and social context matters. For example, while the introduction of generative AI tools gives more of a boost to less-skilled workers, this equalizing force could be a way for workers to increase their earnings potential if compensation is tied to capability. Instead, if firms exploit the higher interchangeability between workers ("why hire an expert copywriter if a less-skilled writer with an AI chatbot can do the same level of work?") these wage gains may never be realized.

AI will likely have outsized impacts on US workers with Bachelors' or Associates' degrees, compared to higher or lower levels of education (Septiandri et al., 2023). This effect could compound over time: if generative AI tools commodify expertise and reduce the returns to specialized skills, workers may no longer spend the time or resources to acquire greater levels of expertise, leading to lower levels of worker skill and overreliance on outsourcing to generative tools. These effects could cause greater competition at the (now larger) lower end of the skill distribution, further depressing wages. There could be further downsides to productivity if non-automatable job tasks would benefit from workers having acquired the sort of foundational knowledge that is now disincentivized.

Governments may play an important role in mitigating the risk of increased inequality and maximizing the productivity potential of new generative AI tools. Explicit policy suggestions are postponed to the "Policymaking in the age of artificial intelligence" section (see also Table 5). Table 2 reports a summary of the main research directions on the impacts of generative AI in the workplace, along with a set of specific example questions. See SM, Table S2, for an extended version.

| **Future research directions** | | | |
| --- | --- | --- | --- |
| **Research area** | **Specific question** | **Potential design** | **Trade-off** |
| Investigate how AI can be designed and implemented to augment human skills and increase productivity, rather than to simply replace workers and forego the long-run productivity upsides of maximizing workers' contributions to production. | How do organizations balance automating knowledge tasks versus hiring more knowledge workers for efficiency and innovation? | Analyze organizational decisions, assessing the ratio of investment in automation technologies to the recruitment of knowledge workers. | Automation offers cost efficiency and 24/7 productivity, but the unique insights brought by human knowledge workers present a compelling case for augmentation over full automation. |
| Examine how AI can facilitate more access to economic opportunities, particularly through reducing language-related barriers and promoting remote work technologies that can democratize access to the digital economy. | What role does AI play in enabling marginalized groups, including countries in the Global South, to access remote work opportunities in the digital economy? | Analyze employment trends and outcomes in marginalized groups before and after introducing AI-enabled remote work platforms. | AI could provide anonymized job matching, enhancing access, but restrictions on technology use could negate these benefits. |
| Conduct long-term studies to monitor the evolving impact of AI on the workforce, capturing both the immediate and delayed effects on work across | How does continuous use of AI tools impact job satisfaction and employee burnout? | Track job satisfaction and burnout rates among employees using AI tools versus those using traditional methods over multiple years. | AI tools might boost innovation and creativity, but could also lead to higher expectations and workload, decreasing satisfaction and increasing burnout. |
| educational and occupational strata. | | | |
| Explore how AI can be utilized in educational and training programs to encourage basic competency with generative AI tools and better-equip workers in vulnerable job sectors in anticipation of labor market changes. | How does intergenerational collaboration within teams, specifically focused to the exchange of AI tool knowledge, impact team performance, inclusivity perceptions, and psychological well-being? | Create diverse teams to assess the impact of intergenerational AI knowledge-sharing on performance, inclusivity perception, and well-being. | Intergenerational knowledge sharing might improve team performance, inclusivity perceptions, and well-being, but it could also introduce challenges such as resistance to change from senior members or frustration among juniors, with negative effects on the outcome variables. |
| Research labor laws, taxation policies, and social support systems that could support workers displaced or disadvantaged by AI. | How do worker retraining and job replacement programs impact the economic stability of workers displaced by AI? | Incorporate AI-skills into pilot worker retraining programs in regions with high rates of AI-driven displacement, comparing against control groups in similar regions with alternative, non-AI worker retraining. | AI-centric worker retraining could enable workers to be more agile in finding re-employment in the event of job displacement, but this may not be successful for all workers, particularly those later in their careers or with fewer pre-existing technical skills. |

*Table 2. Summary of the main research directions on the impact of generative AI in workplace environments, along with specific example questions and corresponding experimental design and theoretical trade-offs.*

# Impact on education

The integration of generative AI in education represents a continuation of the technological evolution that began with Massive Open Online Courses and similar initiatives. Massive courses have increased access to education to some degree (Adil et al., 2024). However, they have often fallen short of their anticipated transformative impact for various reasons, including minimum learning by doing and lack of personal support (Rai & Chunrao, 2016). Generative AI brings a distinctively novel element to educational technology: the role of chatbot tutors, which interact with students to foster skills ranging from prompt engineering to critical thinking and creative ideation. This shift towards using technology as a dynamic partner paves the way to truly skill-adaptive and personalized teaching and on-demand student guidance and support that does not require continuous, intensive investment from educators to repackage content to best meet students' needs. These uses could be particularly effective in large class settings, with significant opportunity to scale-up implementation beyond the capabilities of traditional educational practices. Consequently, generative AI could bridge complex and persistent educational gaps.

A review of AI applications in education identified several use cases that produced higher test scores when students used personalized learning systems (Akgun & Greenhow, 2022). These systems, unlike traditional approaches like static worksheets with standardized questions, detect areas where students lack foundational understanding by adapting educational resources and tools to foster their development. Furthermore, assessment algorithms can expedite grading of written assessments, which supports students by offering timely feedback that can be applied immediately. Students themselves perceive AI as potentially beneficial to their education. College students reported that generative AI provided personalized learning, supported their writing and brainstorming, and assisted with research and analysis (Chan & Hu, 2023). However, students also expressed concerns about the accuracy, privacy, and ethical implications of generative AI tools—including how this technology could adversely impact their personal development and career prospects.

Educational uses of generative AI pose several other challenges. One is the perpetuation of biases and discrimination, potentially reinforcing racial or gender-based stereotypes during personalized learning, automated scoring, and admission processes (Akgun and Greenhow, 2022; Baker and Hawn, 2022). The data used to train AI models could suffer from bias, if those data are based on past human decision making (a notoriously biased process). An example is the translation bias observed in tools like Google Translate, where gender stereotypes are inadvertently perpetuated in language translations. Translating the phrase "she/he is a nurse" from Turkish (which is "genderless") to English (which is "gendered") yielded the feminine form (i.e., "she is a nurse"), while the phrase "she/he is a doctor" yielded the masculine form (i.e., "he is a doctor"; Johnson, 2021). Failing to account for these biases could amplify inequalities and injustices, specifically towards historically marginalized groups.

Given that human educators are susceptible to biases and discrimination, AI systems offer a theoretical advantage: they could be engineered to exhibit less bias. A significant benefit of AI is the ability to audit and address biases within educational systems, a process that proves difficult, if not impossible, with human biases. However, simply introducing slightly less discriminatory technologies into classrooms is not a substitute for the goal of removing discrimination from school (Pasquale, 2020). Moreover, AI systems should be designed with sufficient transparency for users to monitor for and identify potential biases to ensure that these tools effectively serve their intended purposes and reflect the interests of key stakeholders, including students, teachers, and parents (Stoyanovich et al., 2020).

Group-based inequalities may widen because of varying levels of engagement with generative AI tools. For instance, a study revealed that female students report using ChatGPT less frequently than their male counterparts (Carvajal et al., 2024). This disparity in technology usage could not only have immediate effects on academic achievement, but also contribute to future gender gap in the workforce. Therefore, efforts should be made to ensure the benefits of generative AI tools are fairly distributed across all student segments.

It is unclear if Generative AI will alleviate or place increased burdens on teachers. In contrast to the idea that AI tools relieve teachers of repetitive and onerous work, there is growing concern that teachers must engage in additional tasks "behind the scenes" (e.g., curating and filtering content, monitoring student-AI interactions, providing technical support) to ensure that AI tools are able to function in complex classroom settings (Selwyn et al., 2023). This could exacerbate a generational divide among educators, as younger teachers may be more adept with new technology than older teachers. Furthermore, there could be unintended consequences of generative AI on student learning—for example, if students become overly reliant on support from AI tools, this could undermine their capacity to work or think independently. Questions also arise about the accuracy of AI-generated content and the new skills that students must acquire to work effectively with AI systems, such as the ability to evaluate AI-generated content.

The current debate about the role of generative AI, from primary schools to universities, revolves around whether generative AI should be banned, permitted under only some cases, or allowed as assistance for teachers and students. For instance, the New York City education department and Chinese universities have banned generative AI (Elsen-Rooney, 2023; Liu et al., 2023), while the Berlin universities recommended its use in certain scenarios. A growing literature recommends the use of generative AI for teacher and student assistance within the traditional curricula (e.g., Chiu, 2023).

We argue that these approaches are limited in vision. A more forward-thinking approach would involve a curricular revolution to redefine the skills and competencies necessary to effectively utilize generative AI. Calculators did not remove the need for students to learn the properties of algebra and develop mathematical reasoning. Similarly, the internet did not eliminate the need for careful research and fact-checking; in fact, it increased this need, as online information is frequently incorrect or incomplete. In the same vein, generative AI will not eliminate the need to learn effective thought organization, writing, and critical thinking skills. Therefore, curricula must teach how to successfully describe and share ideas, both with and without assistance from generative AI. In addition, they need to emphasize the development of critical-thinking skills, fact-checking abilities, an understanding of how generative AI tools function, and appropriate rules of interaction—including by refraining from anthropomorphizing (and thus misunderstanding) these tools (Kasneci et al., 2023).

More specifically, the text-production abilities of generative AI present an opportunity to teach students critical thinking. This will enable them to evaluate the argument and structure of the generated text and to write intelligent prompts for generative AI. This skill should be recognized and assessed by educators. The output of generative AI is much more variable than other educational technologies (e.g., calculators); therefore, developing these critical thinking abilities and prompt-engineering skills is fundamental.

Another crucial skill is the ability to fact-check generative AI outputs. Fact-checking skills are not taught sufficiently in schools. For instance, among more than 3,000 U.S. high school students and undergraduates, 96 percent did not know how to evaluate the trustworthiness of websites (Breakstone et al., 2021). These fact-checking abilities include smart heuristics such as lateral reading; i.e., the practice of navigating away from an unfamiliar website to verify the reliability of its information by consulting other external sources (McGrew, 2024). A toolbox of similar fact-checking heuristics needs to be developed or remediated for AI-generated content.

Lastly, understanding the nature of large language models, which are statistical machines that calculate correlations between words, is essential. Only in this way can students understand the potential and limits of generative AI, rather than assuming that contemporary generative AI can "think" or "comprehend" like humans.

The adaptation of curricula is challenging, but essential for the future of education. Without such changes, teachers and students may use generative AI merely as an automated assistance tool. This would forego the opportunity to develop higher-order cognitive skills, such as critical judgment and fact-checking, that generative AI itself cannot reliably perform. The result would be a likely decline in higher-order cognitive skills, especially in segments of the population that will use these tools in a more mechanical, less analytical manner. The role of governments in integrating generative AI into the education sector is crucial. We will discuss potential policy recommendations in the final section (see Table 5). Table 3 summarizes the main research directions, along with specific research questions. We refer to SM, Table S3, for an extended list of research questions.

| **Future research directions** | | | |
| --- | --- | --- | --- |
| **Research area** | **Specific question** | **Potential design** | **Trade-off** |
| Examine how generative AI can be effectively used for personalized learning. | How does the use of generative AI for personalized feedback in essay writing impact students' writing proficiency over a semester? | Compare students in classes receiving personalized AI feedback with those receiving standard teacher feedback. Assess improvements in writing proficiency. | Personalized AI feedback could offer more tailored and immediate improvements, but may lack the psychological understanding and motivational impact of human feedback. |
| Investigate how curricula can be redesigned to include generative AI as a tool for enhancing learning while also teaching students to critically engage with and understand this technology. | How does integrating generative AI in science curricula affect students' understanding of complex concepts? | Implement generative AI in a subset of science classes, comparing students' performance on concept understanding to those in classes without AI tools. | Generative AI may enhance concept understanding through personalized learning, but the potential for over-reliance on AI could hinder independent critical thinking skills. |
| Study effective training methods for teachers to integrate AI tools into their teaching practices and identify the additional support required to manage these technologies in the classroom. | Can peer mentoring programs increase teachers' confidence in using generative AI tools? | Implement a peer mentoring program where AI-experienced teachers mentor those less familiar, and assess changes in confidence and usage rates of AI tools in teaching. | Peer mentoring could provide useful support and encouragement, but mismatches in teaching styles might limit the program's effectiveness and even reduce confidence. |
| Explore strategies to ensure that use of generative AI leads to a diversity of | How does the educational use of generative AI affect students' creativity? | Study three groups of students: one using generative AI with educator support for creativity, one using | Generative AI can expand students' creativity with proper guidance but might reduce originality and |

| educational experiences and outcomes | | AI independently, and the other not using AI. Assess the creativity of their outputs. | insight if used without educational support. |
|---|---|---|---|
| Evaluate the long-term impacts of generative AI on student learning, teacher workloads, and educational outcomes. | Does the use of generative AI in lesson planning reduce teachers' preparation time? | Track lesson preparation time for teachers using generative AI tools versus traditional planning methods over a semester. | AI tools could streamline the lesson planning process, but the initial learning curve and adjustments to AI suggestion might offset time savings. |

*Table 3. Summary of the main research directions on the impact of generative AI on education, along with specific research questions, with corresponding experimental design and theoretical trade-offs.*

# Impact on healthcare

Recent advances in AI techniques can democratize healthcare by making efficacious medical care more accessible and affordable. This is often achieved via augmenting human capacities and reducing workload. AI can support clinicians with diagnosis, screening, prognosis, and triaging, alleviating the burden on health practitioners and giving them the "gift of time" (Topol, 2019). For instance, a review of workplace burnout among healthcare providers identified electronic health record systems as a cause of increased stress due to insufficient documentation time, a high volume of patient communications, and negative perceptions by providers (del Carmen et al., 2019). In response, generative AI models may aid in the completion of electronic health record-related tasks, reducing healthcare professionals' administrative demands (Patel and Lam, 2023).

Using AI-systems as clinician co-pilots could also improve diagnostic accuracy and potentially curb biases. For instance, in a randomized test intervention, physicians answered questions around triage, risk, and treatment in chest pain evaluation scenarios, and then reconsidered these answers after receiving advice generated by GPT-4 (Goh et al., 2023). Not only were clinicians willing to heed the advice of the generative AI chatbot, but doing so was also associated with improved accuracy of diagnoses and a reduction in gender and race bias in decisions.

AI systems can also aid in "medical visual question answering"—analyzing medical images (like X-rays or MRI scans) and providing answers to specific questions about these images, typically by leveraging advanced image recognition and AI algorithms (Ren and Zhou, 2020). GPT-4 demonstrates reasonable diagnostic accuracy in simple cases and can answer questions on standardized medical exams, though it struggles with diagnostically complex prompts (Kanjee et al., 2023).

AI-systems could also assist healthcare providers by analyzing and interpreting multimodal clinical data (e.g., photos, radiology images, and surgical videos) to provide relevant information to clinicians (The Lancet Regional Health – Europe, 2023). In one study, endoscopists reviewed colonoscopy videos with and without AI assistance. Their decisions were influenced by AI, particularly when its advice was correct. The integration of human and AI judgment led to superior performance compared to either alone, highlighting effective human-AI collaboration dynamics in medical decision-making (Reverberi et al., 2022). Alas, diagnostic performance of some expert physicians may not be improved by AI—and in fact may cause incorrect diagnoses in situations that otherwise would have been correctly assessed (Agarwal et al., 2023). Further research is needed to identify under what circumstances clinicians should or should not heed the advice of generative AI, and for which clinicians and medical contexts these tools enhance or hinder patient outcomes.

Generative AI could also benefit patients. For instance, AI systems could enable patients to manage their health more proactively through applications that patients can access outside of clinical settings. ChatGPT, for instance, has reasonable accuracy in answering common myths about cancer (Johnson et al., 2023). Research found that people trusted ChatGPT's answers to low-risk medical questions, though trust reportedly varied for questions with greater medical complexity (Nov et al., 2023). Furthermore, ChatGPT's answers to medical questions posted on Reddit's r/AskDocs were rated as higher quality and more empathetic than those of physicians 79% of the time (Ayers et al., 2023).

Conversational agents based on generative AI can also provide greater access to medical advice and simplify medical jargon. This may have positive downstream effects on inequality. Being part of a stigmatized group affects people's engagement and utilization of healthcare services. For example, when contextual cues made racial stereotypes salient, Black women were more likely to feel anxious in a healthcare setting than their White counterparts (Abdou and Fingerhut, 2014). Health professionals are also biased in their treatment of higher-weight patients (Rathbone et al., 2020). As a result, members of stigmatized minority groups are less likely to listen to, or trust, doctors who they perceive as outgroup members (Dovidio et al., 2008). It is possible that interactions between members of stigmatized groups and the healthcare system might be more positive when some decisions are AI-mediated because a patient's stigmatized status may not be as salient as it would be in human-to-human interaction. This suggests that members of stigmatized groups could become more likely to engage with AI-led healthcare because they worry less about group- or identity-based factors affecting their treatment options.

However, there are several important steps required to ensure that this promise comes to fruition. Pre-existing societal biases are often baked into healthcare data—especially when that data consists of human clinicians' decisions and the socioeconomic factors that influence patients' presentation to healthcare facilities. This can make human biases difficult for an AI program trained on that data to overcome. Furthermore, many of the diagnostic criteria and treatment algorithms used in healthcare are also subject to bias, which can drive unequal health outcomes for underserved populations. Although these biases are pervasive in training datasets, there is cause for optimism. Pierson et al. (2021) proposed a model, trained on knee X-rays, that is nearly five times as effective at predicting knee pain in osteoarthritic patients, compared to the traditional Kellgren-Lawrence grading system, which overlooks racial disparities. Obermeyer et al. (2019) found that a healthcare resource-allocation algorithm favored White over Black patients with the same health risk due to its cost-based criteria, but adjusting the focus to patient health eliminated this racial bias.

Such promises in overcoming dataset issues will require updates to laws and regulations governing healthcare data as well as the implementation of incentives to shift the culture of healthcare facilities to promote interoperability and data sharing. While hospitals are developing siloed versions of generative models to facilitate care while preserving privacy, the true potential of AI models will likely rely on methods that leverage the value of large, generalizable datasets.

Benefits aside, patients, medical providers, and those managing healthcare systems may be hesitant to adopt AI due to several psychological barriers. In fact, the impact of AI on clinical practice has been limited despite the growing number of AI tools (Aristidou et al., 2022). One key factor is public trust in AI technologies in healthcare (Quinn et al., 2021). For instance, patients may resist adoption because of misperceptions about AI, such as the belief that AI cannot account for a person's uniqueness as well as a human doctor (Longoni et al., 2019), or because of difficulty in holding AI accountable for mistakes (Promberger and Baron, 2006).

Another factor implicated in adoption hesitancy is the contrast between AI's opaqueness and the illusory perception that human decision-making is more transparent than AI. Decisions made by human physicians or AI are probably equally unobservable to a patient—but because patients feel that they can understand the decision-making as explained by human providers, they ultimately penalize and resist the clinical use of AI (Cadario et al., 2021). The most recent versions of AI tools may be less susceptible to concerns about AI's inscrutability, since the iterative nature of newer generative AI tools may allow patients to ask follow-up questions in a more familiar, conversational format. It is possible that the back-and-forth supported by modern generative AI tools will empower patients with greater information about AI-driven decision-making, at which point patients may be better-equipped to decide whether to trust (or not trust) AI-generated medical recommendations.

Other challenges to AI adoption include pushback from healthcare practitioners—who wish to ensure high quality, experience-driven patient care, or who fear being replaced by machines—and from those managing healthcare systems, who might be reluctant to initiate costly and systemic changes until the usefulness of AI-integration is fully proven.

Insurance markets will also be impacted. Insurers could use AI to refine their practices, capturing a larger share of the surplus. This could lead to welfare losses for consumers. Today, it is not possible to determine highly accurate, individualized probabilities for the future health conditions of a particular insurant—insurance as a field relies instead on population-level probabilities, with some refinement from explicit risk factors. However, if generative AI allows companies to more accurately estimate this probability—for example, by incorporating information from unobservable factors that are identifiable only through advanced machine learning algorithms run on text-based claims, electronic health records, or other data—they might charge higher premiums to those at greater risk without offering reductions to those at lower risk. This could remove health care access to high-risk populations in private insurance-based systems. Use of AI is prime for abuse by insurers who wish to maximize profits over ensuring equitable access to healthcare.

AI could also enable insurers to reach currently uninsured groups, reducing inefficiencies and achieving market completeness. A concrete example of this is the use of Responsible Artificial Intelligence in healthcare to predict and prevent insurance claim denials, which could lead to significant cost savings and improved patient well-being (Johnson et al., 2021). Moreover, the application of AI by insurance companies might allow for a more accurate prediction of loss probabilities, thus reducing one of the industry's most inherent problems, namely asymmetric information (Eling et al., 2022).

Generative AI may come to fulfill social needs for some people, which could have downstream effects on health. There is robust evidence linking social connectedness or lack thereof to long-term health outcomes (Holt-Lunstad, 2021), including increased risk for chronic illnesses such as cardiovascular disease and stroke (Valtorta et al., 2016), type 2 diabetes, and dementia (Penninkilampi et al., 2018), as well as mortality from all causes (Holt-Lunstad et al., 2010). While digitally mediated forms of socializing (e.g., social media) have been utilized for years, there is increasing concern about the implications of these platforms for mental, social, and physical health (Haidt, 2024). Generative AI can be used as a conversational companion, potentially replacing some human interactions. Indeed, digital proxies for social connection may, with increasing sophistication, mimic features of social connection, which could in turn decrease motivation to develop authentic human relationships. These features may relieve some of the tensions of human connection, leading people to preferentially spend more time with AI than humans or even form pseudo-social attachments to AI systems.

If AI-based chatbots are insufficient stand-ins for customary human interactions (which is likely true), then many negative consequences could result. Humans are social beings, so our biological systems can become dysregulated when social needs are unmet, leading to poorer health (Beckes and Sbarra, 2022). Therefore, it is essential that some key elements of customary human interactions be retained. For example, research finds that relative to emails and other text-based interactions, those involving human voice boost social connection (Kumar and Epley, 2021). At the same time, AI-based chatbots could be useful to add social experiences for some individuals (while not completely replacing human-to-human interaction), particularly for those facing difficulties developing relationships on their own (who need "Vitamin S," from Social contact; Van Lange and Columbus, 2021), but are likely to be a poor or even dangerous replacement for human interaction writ large.

In sum, generative AI presents significant opportunities to alleviate inequalities in physical and mental health, in addition to augmenting healthcare providers' capabilities. However, it is crucial to ensure that generative AI are only designed to supplement, rather than replace, human social interactions. Excessive dependence on AI for social engagement could lead to various adverse outcomes, including social isolation and deteriorating mental and physical health. Table 4 outlines key areas for future research, along with specific example questions, potential designs, and theoretical trade-offs. See SM, Table S4, for an extended list of research questions.

| Future research directions | | | |
| --- | --- | --- | --- |
| **Research area** | **Specific question** | **Potential design** | **Trade-off** |
| Research how AI can assist healthcare professionals in diagnosis, treatment planning, and patient monitoring. | Does AI-assisted medical decision making improve patient outcomes including decision accuracy and bias reduction? | Healthcare providers are given the option to consult or not consult medical advice from generative AI chatbots, with decision accuracy and decision bias measured across medical contexts and healthcare provider characteristics. | Clinicians could heed the advice of AI, improving their medical decision making. However, mistrust for AI tools, low AI literacy, or other sources of suboptimal interactions with AI could render AI-assisted decision making ineffective or detrimental for patient outcomes. |
| Investigate the use of AI to reduce the administrative burden on healthcare providers through efficient electronic health records (EHR) management. | Does the integration of AI into EHR systems reduce the time healthcare providers spend on documentation compared to traditional EHR systems? | Healthcare providers are divided into two groups, one using AI-integrated EHR systems and the other using traditional EHR systems, with the time spent on documentation measured over a specified period. | AI could streamline documentation processes, reducing time spent on paperwork. However, learning curves associated with new AI systems might initially increase the time required for documentation. |
| Study how AI can contribute to the development of personalized medicine, adapting treatments to individual patient needs and reducing | Can AI-based predictive models accurately identify patients at high risk for diabetes (or any other disease) and guide early | Develop AI models using patient health data to predict diabetes risk and apply intervention strategies based on model predictions. Compare the | Predictive models could allow for earlier and more targeted interventions. However, inaccuracies in predictions could |

| healthcare disparities. | intervention strategies? | incidence of diabetes in this group with a control group receiving standard care. | lead to unnecessary interventions or miss at-risk individuals. |
|---|---|---|---|
| Investigate strategies to increase public trust and understanding of AI in healthcare. | Does providing patients with detailed explanations of AI diagnostic processes improve their trust in AI-driven healthcare? | Patients receiving AI-driven diagnostics are split into two groups, one receiving detailed explanations and the other receiving standard information. Trust levels are measured post-interaction. | Detailed explanations could demystify AI processes, increasing trust. However, overly technical or complex information might overwhelm patients, potentially reducing trust. |
| Research how AI can improve healthcare accessibility in underserved regions and populations, in both rural and urban areas. | Can AI-driven mobile health applications effectively increase healthcare access in rural communities? | Implement an AI-driven mobile health application in rural communities. Compare health outcomes, access to care, and self-management behaviors with a control group not using the application. | Mobile health apps could significantly improve access and self-management, but poor internet connectivity and low digital literacy in rural areas might limit effectiveness and even widen disparities. |
| Investigate the potential of AI to facilitate social connections, particularly for individuals with difficulties in forming relationships, while also studying the potential risks of over-reliance on AI for social interaction. | Can AI-based chatbots effectively reduce feelings of loneliness compared to traditional social programs? | Participants are divided into two groups, with one interacting with AI-based chatbots and the other engaging in traditional social programs. Measures of loneliness and social skills are assessed over time and their influence on health. | AI agents could offer constant companionship, potentially reducing loneliness. However, overreliance on AI could inhibit motivation for human interaction leading to increased isolation. Further, the lack of genuine human interaction might not fulfill deep social needs, |

|  |  |  | affecting overall health. |
|---|---|---|---|

*Table 4. Summary of the main research directions on the impact of generative AI on healthcare, along with specific example questions.*

# Policymaking in the age of artificial intelligence

**Regulation of AI**

The rapid popularization of generative AI models has prompted many governments worldwide to begin building regulatory frameworks. The challenges raised by generative AI are global in nature (Jobin et al., 2019). However, the responses to these challenges so far have been specific to individual countries or areas. In this article, we focus on the regulatory responses of the EU, US, and UK. Regulations are also being developed in other major countries, including China and India (Haridas et al., 2023; Roberts, 2023).

The European Union's AI Act has emerged as one of the first major attempts to provide a legal framework for the development and deployment of AI. The Act aims to address the challenges posed by AI technologies while fostering innovation and trust in AI applications (European Parliament, 2023). This initiative comes with several pros. Firstly, it introduces a risk-based regulatory approach, distinguishing among banned, high-risk, and low-risk AI applications. This categorization ensures that AI systems with significant implications for individuals' rights and safety are subject to stricter scrutiny and compliance requirements. Secondly, the Act emphasizes transparency and accountability in AI systems, requiring clear information about how AI decisions are made, particularly in high-risk scenarios. Additionally, the Act promotes ethical AI development, focusing on fundamental rights, non-discrimination, and privacy. However, the Act is not without its cons (Morgan Lewis, 2023). The broad definitions and categories within the Act pose challenges, creating potential uncertainty for AI developers and users. Further, the strict regulations might place EU companies at a competitive disadvantage globally, particularly against firms in regions with more lenient AI laws.

In contrast, the US has historically had a more fragmented approach, with various federal and state-level initiatives rather than a single, comprehensive legislative framework (Felz et al., 2022). This approach has its advantages. For one, it allows for more flexibility and adaptability in regulation, catering to the diverse range of AI applications and industries in the US. It also promotes a more innovation-friendly environment by avoiding overly prescriptive rules that could hinder technological advancement. However, it also has notable disadvantages. The lack of a unified regulatory framework can lead to inconsistencies and uncertainties, potentially creating a complex patchwork of regulations for AI companies to navigate. This fragmented approach might also lag in addressing broader ethical and social concerns about AI, such as privacy, bias, and accountability. Further, without a cohesive national strategy, the US risks falling behind in setting global standards for AI governance. On October 30, 2023, President Biden issued an Executive Order on Safe, Secure, and Trustworthy Artificial Intelligence, which directs the development of new guidelines, reports, and governance structures relating to AI, representing an effort to establish a more cohesive federal policy on AI (White House, 2023).

In the UK, the government has published a White Paper advocating for a pro-innovation approach, particularly in commercial applications of AI (Department for Science Innovation and Technology, 2023). While the White Paper recognizes the risks of AI and the challenge of building public trust, it refrains from proposing a regulatory framework to encourage innovation. Instead, the White Paper outlines some "cross-sectoral" non-statutory soft principles: safety, security, robustness, appropriate transparency and explainability, fairness, accountability and governance, and contestability and redress. The White Paper opts against a specialist AI regulator, preferring to support existing regulators in integrating AI considerations. Furthermore, the focus on commercial innovations has drawn criticism for overlooking the increasing use of AI in government sectors like healthcare and education. One leading non-governmental organization, the Public Law Project, led a civil society coalition to produce *Key principles for an alternative AI white paper* (Public Law Project, 2023), which argues that an alternative vision is necessary. Amongst other proposals, the alternative white paper argues that: government use of AI must be transparent, transparency requirements must be mandatory, there must be clear mechanisms for accountability, the public should be consulted about new automated decision-making tools before they are deployed by government, there must be a specialist regulator to enforce the regulatory regime and ensure people can seek redress when things go wrong, and uses of AI that threaten fundamental rights should be prohibited.

We believe that the regulations of the EU, the US, and the UK do not pay sufficient attention to socioeconomic inequalities. In the following, we outline several key interventions currently missing from these regulatory frameworks (Acemoglu, Autor et al., 2023). See Table 5 for a summary.

**Tax system:** It has long been recognized that tax codes in many developed countries often place a heavier burden on firms that hire labor than on those that invest in algorithms to automate work (Abbott and Bogenschneider, 2018; Acemoglu et al., 2020). This has resulted in a lower share of income to labor while capital investments are rewarded. We should aim to create a more symmetric tax structure, where marginal taxes for hiring (and training) labor and for investing in equipment/software are equated. This would help to shift incentives toward human-complementary technological choices by reducing the bias of the tax code toward physical capital over human capital. As Bill Gates declared back in 2017, "If a robot comes in to do the same thing, you'd think that we'd tax the robot at a similar level" (Quartz, 2017).

**Labor voice and control of consumer information:** Given that AI will have tremendous impact across industries and throughout society, it would be prudent to ensure that workers and civil society have a voice in this change. Health and safety rules should also be updated accordingly. In addition, data unions could be helpful to put the power and benefits of user data back in the hands of consumers. Given the concerns that a handful of very large companies will control the direction of generative AI, it is reasonable that users be compensated for the use of their information, or enabled to support other emergent competitors to predominant market players like Microsoft and Google.

**Funding for more human-complementary research:** Because the current path of research is biased toward automation (Acemoglu and Restrepo, 2019; Acemoglu and Johnson, 2023), support for research and development of human-complementary AI technologies could offer strong upsides for growth. It is most feasible to focus on specific sectors and activities where opportunities are already abundant. These include education, healthcare, and modern craft worker training—where the information provisional capabilities of AI systems could boost productivity and enable workers to earn higher wages by augmenting their skills. In the US, DARPA orchestrated investments and competitions to foster development of self-driving cars and dexterous robotics—in a similar fashion, governments should encourage competition and investment that pairs AI tools with human expertise, aiming to improve work in vital social sectors.

**Professional development and training:** Investment in professional development and training is crucial for professionals such as educators and healthcare workers to effectively integrate AI tools into their work. Training programs should focus on the capabilities and limitations of AI, include ethical considerations, and teach technical skills required to interact with AI systems. Such training will empower professionals to use AI as a complementary tool that enhances their skills.

**Combating AI-generated misinformation:** Given the substantial impact that generative AI can have on misinformation, it is critical for governments to invest in combating AI-generated misinformation. Tools and standards to identify AI-generated content, including text, images, audio, and video, should be developed. Additionally, educational campaigns should be initiated, to reduce general susceptibility to misinformation and provide the public with improved fact-checking strategies. A task force composed of policymakers, technology companies, and social scientists could help develop practical methods to effectively combat a potential infodemic.

**Governmental and consultative expertise:** To foster human-complementary AI integration, it is fundamental to have AI expertise within the government. AI will touch every area of government investment, regulation, and oversight. Developing consultative AI bodies that can advise governments and support the many agencies and regulators tackling these challenges will support more timely and effective decision-making. Initiatives like the High-Level Expert Group on Artificial Intelligence, established by the EU, the Responsible Technology Adoption Unit, set by the UK, and the Artificial Intelligence topic in the US National Institute of Standards and Technology represent significant progresses in this direction.

| **Policy recommendations** |
|---|
| Create a more symmetric tax structure, where marginal taxes for hiring and training labor and for investing in the development, installation and usage of new AI tools are equated. |
| Involve workers and civil society in AI-related changes and establish data unions to empower data owners with control over their data. |
| Increase support for research into human-complementary AI tools to enhance productivity and workers' skillsets. |
| Train professionals, especially in education and healthcare, in the use of AI tools, covering their capabilities, limitations, and ethical considerations. |
| Invest in strategies aimed to combat AI-generated misinformation, including developing tools that can identify AI-generated misinformation and initiating educational campaigns. |
| Embed AI expertise within government to advise and support decision-making across various sectors. |

*Table 5. Policy recommendations for mitigating socioeconomic inequalities potentially caused by generative AI in the workplace, education, healthcare, and information, and not covered by current regulatory approach in the EU, US, and UK.*

**Regulation using AI**

Generative AI holds enormous potential to provide policy suggestions, due to its capacity to analyze vast amounts of data, recognize complex patterns, and offer insights that might elude human analysis. Such analysis can uncover hidden relationships and forecast future trends, providing a data-driven foundation for policy decisions. Moreover, AI's potential to simulate various policy outcomes based on historical data and predictive models can aid policymakers in understanding the potential impacts of their decisions (Zheng et al., 2022; Madan and Ashok, 2023). However, the ethical and practical concerns of using AI for policymaking are significant and perhaps even prohibitive with the current tools available.

For AI-powered systems to be reliable decision-making assistants, they must be able to understand and complement human behavior in decision-making contexts. An emerging line of research has begun investigating how chatbots powered by large language models make decisions when asked to behave like humans in various contexts, including cooperative, altruistic, trust, moral, risk, time, and food scenarios (Chen et al., 2023; Mei et al., 2024). In synthesis, these findings suggest that chatbots behavior is similar to human behavior, leading to the suggestion that "GPT could have the potential in assisting human decision-making" (Chen et al., 2023). Furthermore, sentiment scores provided by GPT-4 have been shown to explain how humans balance self-interest and the interest of others, beyond economic models based solely on monetary outcomes (Capraro et al., 2024). While this research highlights the ability of large language models to simulate human behavior, it has also been observed that GPT-4 consistently underestimates people's self-interest and inequity-aversion, while overestimating their level of altruistic behavior (Capraro et al., 2023). This "optimistic bias" carries important consequences for those creating and using AI, as assumptions of excessive human prosociality could result in disillusionment, frustration, and even social tensions (Capraro et al., 2023).

Additionally, the alignment between AI-powered decision-making assistants and humans will inevitably become more complex in situations with multiple and potentially competing values. Humans have a wide range of culturally diverse beliefs about right and wrong, and aligning AI systems to human values and preferences is challenging even within narrow domains such as automated driving (Awad et al., 2018). Aligning generative AI, especially in the domain of policy recommendations, becomes even more challenging. For example, consider different normative principles that have been identified for trustworthy ethical AI (Floridi and Cowls, 2019). It is argued that AI should, amongst other things, promote beneficence (promote human well-being and welfare); non-maleficence (not cause harm and generate outputs that assist in carrying out illegal, harmful, or immoral actions); justice (preserve fairness, justice, and solidarity: it should not generate outputs that discriminate against certain groups, especially marginalized groups), and ensure autonomy (respecting human freedom and ensuring humans should choose how and whether to delegate policy decisions). While these principles are all defensible in the abstract—forming the basis of much normative ethical theory and applied ethics—challenges will inevitably arise when these principles conflict. Generative AI may assist in generating policies that maximize overall aggregate welfare, in line with utilitarian philosophy (Mill, 1861) but in doing so infringe on human rights (neglecting the principle of autonomy) or recommending some harm to a smaller group for the benefit of the majority (neglecting the principle of non-maleficence). There is no consensus amongst laypeople about how such moral dilemmas should be resolved (Greene et al., 2001; Kahane et al., 2018), nor is there normative agreement amongst philosophers on how they should be resolved and why. This discord among human thinkers underscores the challenge of programming AI to make policy decisions that involve moral trade-offs.

To make an explicit example, consider the following three high-level, high-priority constraints for aligned chatbots: 1) to not cause harm or provide dangerous information; 2) to not generate outputs that discriminate against certain groups, and 3) to be culturally sensitive. While these objectives are all desirable, they are increasingly difficult to reconcile. If the only constraint is to avoid dangerous information (non-maleficence), regardless of social neutrality or cultural sensitivity (justice), one can use reinforcement learning from human feedback using a convenience sample of annotators. But this approach would fail to ensure social neutrality, since a convenience sample of annotators would have unrepresentative biased views on what constitutes immoral actions or generally undesirable outputs (Peters, 2022). Political and social neutrality may be approximated by engaging in carefully balanced reinforcement learning from human feedback, based on a broadly representative array of opinion, or by having a singular chatbot that facilitates consensus-making among diverse human values (Bakker et al., 2022). Alternatively, an ecosystem of chatbots with diverse systems of values—liberal and conservative bots, secular and religious bots, etc.—may emerge. These chatbots can each focus on their specific domain, while also undergoing a political process to achieve collective decisions among themselves. However, these approaches would still fail at cultural sensitivity, since different cultures may be different in terms of the social groups they include, the topics these groups value, and the range of these cultural values (Atari et al., 2023).

In the worst case scenario, then, the alignment of generative AI would be entirely based on the views of a small group of socially, politically, and culturally homogeneous informants (Acemoglu and Johnson, 2023). But even in the best-case scenario, where generative AI is trained on a diverse and nuanced set of preferences, we would still have significant problems. Even if we have a more diverse set of information about humans' actual values and how they might want trade-offs to be made for moral dilemmas, we still lack widespread agreement on a specific normative standard to justify these descriptive preferences.

Aside from the alignment problem, there is the implementation problem: how to equip policymakers with reliable support from organizations and specialized staff. Most policymakers currently lack the knowledge and skills to directly evaluate the extent to which AI-based generative chatbots may embed undesired preferences or detrimental systematic biases. Admittedly, one can hardly expect that policymakers can acquire the needed knowledge and skills in due time. So, policymakers are likely to become the "principals" in a principal-agent problem, struggling to consider the preferences of their "AI-agents". Policymakers will have to rely on some other agent for this evaluation, based on scientific principles for characterizing machine behavior and misbehavior (Rahwan et al., 2018).

Hence, it is crucial to design supporting organizations that systematically provide the policymakers with: (i) frequent evaluation of the current state of alignment between legal and regulatory requirements, and (ii) mechanisms to signal any legal and regulatory changes in those requirements to companies that use AI-based chatbots—thus putting *society in the loop* (Rahwan, 2018). These desiderata in turn require the construction of a dedicated office in the organization that monitors AI-based chatbots, considering—and possibly predicting—their evolution.

Finally, there is also a philosophical problem. Even if we could solve the problems of conflicting preferences, even if we could generate a good culturally sensitive sample, and even if we could solve the implementation problem - should we?

## Conclusion

The future will likely be starkly different from anything we have experienced before. But the effects of generative AI will ultimately depend on the choices that we make to design and deploy the technology. We stand at a unique and historical moment; our decisions and actions today will shape the trajectory of our future. This responsibility extends to all sectors of society, including governance, scientific research, industry, and the public.

We have focused on the socioeconomic inequalities that are likely to be impacted—for better or worse—by the advent of generative AI. This technology has profound implications in the domain of information, where it has the potential to offer more tailored, efficient, and democratic ways to process information. Yet it also poses several challenges, including anticompetitive market advantages, data misuse and abuse, and misinformation. These mixed outcomes will certainly affect a very wide range of social organization and decision making.

Here, we focused specifically on the domains of work, education, and healthcare. For instance, in the workplace, AI could automate some job tasks, create new work, but also change wage distributions, and require new skill sets. In education, AI could democratize learning and provide personalized education solutions, but also increase the digital divide. In the healthcare sector, AI's ability to analyze large datasets can lead to better patient outcomes, but it also raises questions about equitable access to AI-driven healthcare services and genuine human interactions.

We have outlined several research questions that urgently require answers to address these issues effectively. These questions aim to harness AI's benefits while mitigating its risks. Additionally, we have observed that current regulatory approaches in the EU, US, and UK sometimes fail to adequately address these challenges. There is a need for a dynamic regulatory framework that can keep pace with the rapid advancements in AI technology.

Our hope is that this work contributes to a comprehensive research agenda and public debates on these critical topics. As we noted at the beginning of this paper, The rise of powerful AI will be either the best, or the worst thing, ever to happen to humanity. We may not yet know which, because we do not know how humans will react to this technology. As we stand at the cusp of this new era of human-machine interactions, it is crucial that we engage in thoughtful and inclusive discussions about the role of AI in shaping our society, because the decisions we make today will have lasting impacts on generations to come.

**Acknowledgments**

DA acknowledges support from the Hewlett Foundation, the NSF, Schmidt Sciences, Google, the Sloan Foundation, the Smith Richardson Foundation, and the Washington Center for Equitable Growth. JFB acknowledges support from grant ANR-19-PI3A-0004, grant ANR-17-EURE-0010, and the research foundation TSE-Partnership. PBG acknowledges support from the Spanish Ministry of Science and Innovation (PID2021-126892NB-I00). KD acknowledges support from the European Research Council Advanced Grant "Consequences of conspiracy theories - CONSPIRACY_FX" Number: 101018262. JACE acknowledges support by the ESRC (ES/V015176/1) and Leverhulme Trust (PLP-2021-095). JJ acknowledges support from the Australian Research Council Laureate Fellowship (FL180100094). SJ acknowledges support from the Hewlett Foundation and the Sloan School, MIT. SN acknowledges support from the University of Turin, Italy, Grant "Human-Machine Communication Cultures: Artificial Intelligence, Media and Cultures in a Global Context", reference number NATS_GFI_22_01_F. PVL acknowledges support from an Ammodo science award (2020) provided by the Royal Netherlands Academy of Arts and Sciences. JVB acknowledges support from the Trust in generative AI, Google Jigsaw, and from the Center for Conflict and Cooperation, Templeton World Charity Foundation (TWCF-2022-30561).

# References

Abbott, R., & Bogenschneider, B. (2018). Should Robots Pay Taxes? Tax Policy in the Age of Automation. *Harvard Law & Policy Review, 12*, 145-175.

Abdou, C. M., & Fingerhut, A. W. (2014). Stereotype threat among black and white women in healthcare settings. *Culturally Diverse Ethnic Minority Psychology, 20*, 316–323.

Acemoglu, D. (2002). Technical change, inequality, and the labor market. *Journal of Economic Literature, 40(1),* 7–72.

Acemoglu, D. (2023). Written testimony for hearing on "The Philosophy of AI: Learning from History, Shaping Our Future." *Senate Committee on Homeland Security and Governmental Affairs.* Retrieved from https://www.hsgac.senate.gov/wp-content/uploads/Testimony-Acemoglu-2023-11-08.pdf

Acemoglu, D. (2024). Harms of AI. In J. Bullock (Ed.), *The Oxford Handbook of AI Governance,* forthcoming. Oxford University Press. Retrieved December 4, 2023, from https://academic-oup-com.libproxy.mit.edu/edited-volume/41989/chapter/411053764

Acemoglu, D., Autor, D., Dorn, D., Hanson, G., & Price, B. (2016). Import competition and the great U.S. employment sag of the 2000s. *Journal of Labor Economics, 34*, 141-198.

Acemoglu, D., Autor, D., & Johnson, S. (2023). Can we have pro-worker AI? MIT Shaping the Future of Work Initiative, policy memo. Retrieved from https://shapingwork.mit.edu/wp-content/uploads/2023/09/Pro-Worker-AI-Policy-Memo.pdf

Acemoglu, D., & Johnson, S. (2023). *Power and progress: Our 1000-year struggle over technology and prosperity*. PublicAffairs, Hachette.

Acemoglu, D., Manera, A., & Restrepo, P. (2020). Does the U.S. tax code favor automation? *Brookings Papers on Economic Activity.* Retrieved December 7, 2023, from https://www.brookings.edu/articles/does-the-u-s-tax-code-favor-automation/

Acemoglu, D., Ozdaglar, A., & Siderius, J. (2023). A Model of Online Misinformation. *Review of Economic Studies*, forthcoming.

Acemoglu, D., & Restrepo, P. (2018). The race between man and machine: Implications of technology for growth, factor shares, and employment. *American Economic Review, 108*, 1488-1542.

Acemoglu, D., & Restrepo, P. (2019). Automation and new tasks: How technology displaces and reinstates labor. *Journal of Economic Perspectives, 33*, 3-30.

Acemoglu, D., & Restrepo, P. (2022a). Tasks, automation, and the rise in U.S. wage inequality. *Econometrica, 90*, 1973-2016.

Acemoglu, D., & Restrepo, P. (2022b). Demographics and automation. *Review of Economic Studies, 89*, 1-44.

Acerbi, A., Altay, S., & Mercier, H. (2022). Research note: Fighting misinformation or fighting for information? *Harvard Kennedy School (HKS) Misinformation Review*, *3*.

Adil, H. M., Ali, S., Sultan, M., Ashiq, M., & Rafiq, M. (2024). Open education resources' benefits and challenges in the academic world: a systematic review. *Global Knowledge, Memory and Communication*, *73*, 274-291.

Agarwal, N., Moehring, A., Rajpurkar, P., & Salz, T. (2023). Combining Human Expertise with Artificial Intelligence: Experimental Evidence from Radiology. NBER Working Paper No. 31422.

Akgun, S. & Greenhow, C. (2022). Artificial intelligence in education: Addressing ethical challenges in K-12 settings. *AI Ethics, 2,* 431–440.

Alam, M.F., Lentsch, A., Yu, N., Barmack, S., Kim, S., Acemoglu, D., Hart, A.J., Johnson, S., & Ahmed, F. (2024). From Automation to Augmentation: Redefining Engineering Design and Manufacturing in the Age of NextGen-AI. *An MIT Exploration of Generative AI*. https://mit-genai.pubpub.org/pub/9s6690gd.

Aristidou, A., Jena, R., & Topol, E. J. (2022). Bridging the chasm between AI and clinical implementation. *Lancet, 399*, 620.

Atari, M., Xue, M. J., Park, P. S., Blasi, D. E., & Henrich, J. (2023). Which Humans? *Available at: https://doi.org/10.31234/osf.io/5b26t*

Autor, D. (2019). Work of the Past, Work of the Future. *AEA Papers and Proceedings, 109*, 1–32.

Autor, D. (2024). Applying AI to rebuild middle class jobs. *NBER Working Paper No. 32140.*

Autor, D., Chin, C., Salomons, A., & Seegmiller, B. (2022). New frontiers: The origins and content of new work, 1940–2018. *NBER Working Paper no. 30389*.

Autor, D., Katz, L., & Krueger, A. (1998). Computing inequality: Have computers changed the labor market? *Quarterly Journal of Economics, 113*, 1169–1213.

Awad, E., Dsouza, S., Kim, R., Schulz, J., Henrich, J., Shariff, A., Bonnefon, J. F., & Rahwan, I. (2018). The moral machine experiment. *Nature, 563*, 59-64.

Ayers, J. W. et al (2023). Comparing physician and artificial intelligence chatbot responses to patient questions posted to a public social media forum. *JAMA Internal Medicine, 183*, 589-596.

Baker, R. & Hawn, A. (2022). Algorithmic bias in education. *International Journal of Artificial Intelligence in Education, 32*, 1052–1092.

Bakker, M., Chadwick, M., Sheahan, H., Tessler, M., Campbell-Gillingham, L., Balaguer, J., ... & Summerfield, C. (2022). Fine-tuning language models to find agreement among humans with diverse preferences. *Advances in Neural Information Processing Systems, 35*, 38176-38189.

Beckes, L., & Sbarra, D. A. (2022). Social baseline theory: State of the science and new directions. *Current Opinion in Psychology, 43*, 36-41.

Beller, I. (2023). Tracker: State legislation on deepfakes in elections. *Public Citizen*. https://www.citizen.org/article/tracker-legislation-on-deepfakes-in-elections

Benson, T. (2023). This disinformation is just for you. *Wired*. https://www.wired.com/story/generative-ai-custom-disinformation/

Bostrom, N. (2017). *Superintelligence*. Dunod.

Breakstone, J., Smith, M., Wineburg, S., Rapaport, A., Carle, J., Garland, M., & Saavedra, A. (2021). Students' civic online reasoning: A national portrait. *Educational Researcher*, *50*, 505-515.

Brynjolfsson, E., Li, D., & Raymond, L. (2023). Generative AI at work. *NBER Working Paper No. 31161*.

Brynjolfsson, E., & McAfee, A. (2016). *The second machine age: Work, progress, and prosperity in a time of brilliant technologies*. W.W. Norton.

Bubeck, S., Chandrasekaran, V., Eldan, R., Gehrke, J., Horvitz, E., Kamar, E., ... & Zhang, Y. (2023). Sparks of artificial general intelligence: Early experiments with gpt-4. *Available at: https://arxiv.org/abs/2303.12712*.

Cadario, R., Longoni, C., & Morewedge, C. K. (2021). Understanding, explaining, and utilizing medical artificial intelligence. *Nature Human Behaviour*, *5*, 1636-1642.

Capraro, V., & Celadin, T. (2023). "I think this news is accurate": Endorsing accuracy decreases the sharing of fake news and increases the sharing of real news. *Personality and Social Psychology Bulletin*, *49*, 1635-1645.

Capraro, V., Di Paolo, R., Perc, M., & Pizziol, V. (2024). Language-based game theory in the age of artificial intelligence. *Journal of the Royal Society Interface*, *21*, 20230720.

Capraro, V., Di Paolo, R., & Pizziol, V. (2023). Assessing Large Language Models' ability to predict how humans balance self-interest and the interest of others. *arXiv preprint arXiv:2307.12776*.

Carvajal, D., Franco, C., & Isaksson, S. (2024). Will Artificial Intelligence Get in the Way of Achieving Gender Equality?. *Available at https://openaccess.nhh.no/nhh-xmlui/bitstream/handle/11250/3122396/DP%2003.pdf*

Chen, Y., Liu, T. X., Shan, Y., & Zhong, S. (2023). The Emergence of Economic Rationality of GPT. *Proceedings of the National Academy of Sciences, 120*, e2316205120.

Citron, D. K. & Pasquale, F. A. (2014). The scored society: Due process for automated predictions. *Washington Law Review, 89*, 1-33.

Chan, C. K. Y. & Hu, W. (2023). Students' voices on generative AI: Perceptions, benefits, and challenges in higher education. *International Journal of Educational Technology in Higher Education, 20*(1). https://doi.org/10.1186/s41239-023-00411-8

Chiu, T. K. F. (2023). The impact of Generative AI (GenAI) on practices, policies and research direction in education: A case of ChatGPT and Midjourney. *Interactive Learning Environments*.

Chugunova, M., & Sele, D. (2022). We and It: An interdisciplinary review of the experimental evidence on how humans interact with machines. *Journal of Behavioral and Experimental Economics, 99*, 101897.

Cohn, A., Gesche, T., & Maréchal, M. A. (2022). Honesty in the digital age. *Management Science*, 68, 827-845.

Costello, T. H., Pennycook, G., & Rand, D. G. (2024). Durably reducing conspiracy beliefs through dialogues with AI. *Available at https://osf.io/preprints/psyarxiv/xcwdn*

Counts, S., Suri, S., Brown, A., Xu, B., R Raghavan, S. (2022). Who gets to work in the digital economy? *Business and Society.* Available at: https://hbr.org/2022/08/who-gets-to-work-in-the-digital-economy

Cutler, K. (2023). ChatGPT and search engine optimisation: The future is here. *Applied Marketing Analytics, 9*, 8-22.

Del Carmen, M. G., Herman, J., Rao, S., Hidrue, M. K., Ting, D., Lehrhoff, S. R., ... & Ferris, T. G. (2019). Trends and factors associated with physician burnout at a multispecialty academic faculty practice organization. *JAMA Network Open*, *2*, e190554-e190554.

del Rio-Chanona, M., Laurentsyeva, N., & Wachs, J. (2023). Are Large Language Models a Threat to Digital Public Goods? Evidence from Activity on Stack Overflow. *Available at https://arxiv.org/abs/2307.07367*

Dell'Acqua, F., McFowland, E., Mollick, E. R., Lifshitz-Assaf, H., Kellogg, K., Rajendran, S., ... & Lakhani, K. R. (2023). Navigating the jagged technological frontier: Field experimental evidence of the effects of AI on knowledge worker productivity and quality. *Harvard Business School Technology & Operations Mgt. Unit Working Paper*, (24-013).

Department for Science, Innovation, and Technology (2023). AI-regulation: A pro-innovation approach. https://www.gov.uk/government/publications/ai-regulation-a-pro-innovation-approach

Del Vicario, M., Bessi, A., Zollo, F., & Quattrociocchi, W. (2016). The spreading of misinformation online. *Proceedings of the National Academy of Sciences, 113*. 554-559.

Dobber, T., Metoui, N., Trilling, D., Helberger, N., & de Vreese, C. (2021). Do (microtargeted) deepfakes have real effects on political attitudes?. *The International Journal of Press/Politics*, *26*, 69-91.

Douglas, K. M. (2021). Are conspiracy theories harmless? *The Spanish Journal of Psychology, 21,* e13.

Dwivedi, Y. K., Kshetri, N., Hughes, L., Slade, E. L., Jeyaraj, A., Kar, A. K., ... & Wright, R. (2023). "So what if ChatGPT wrote it?" Multidisciplinary perspectives on opportunities, challenges and implications of generative conversational AI for research, practice and policy. *International Journal of Information Management*, *71*, 102642.

Dovidio, J. F., Penner, L. A., Albrecht, T. L., Norton, W. E., Gaertner, S. L., & Shelton, J. N. (2008). Disparities and distrust: The implications of psychological processes for understanding racial disparities in health and health care. *Social Sciences and Medical Journal, 67*, 478–486.

Ehrenfeld, J.M. (2023). AMA president sounds alarm on national physician shortage. *American Medical Association, press release*. https://www.ama-assn.org/press-center/press-releases/ama-president-sounds-alarm-national-physician-shortage

Eling, M., Nuessle, D. & Staubli, J. (2022). The impact of artificial intelligence along the insurance value chain and on the insurability of risks. *The Geneva Papers on Risk and Insurance: Issues and Practice, 47*, 205–241.

Elsen-Rooney, M. (2023). NYC education department blocks ChatGPT on school devices, networks. https://www.chalkbeat.org/newyork/2023/1/3/23537987/nyc-schools-ban-chatgpt-writing-artificial-intelligence/

European Parliament (2023). https://www.europarl.europa.eu/news/en/press-room/20230505IPR84904/ai-act-a-step-closer-to-the-first-rules-on-artificial-intelligence

Felz, D. J., Peretti, K. K., & Austin, A. (2022). Privacy, cyber and data strategy advisor: AI regulation in the U.S.: What's coming, and what companies need to do in 2023. https://www.alston.com/en/insights/publications/2022/12/ai-regulation-in-the-us

Ferrara, E., Varol, O. Davis, C., Menczer, F. & Flammini, A. (2016). The rise of social bots. *Communications of the Association for Computing Machinery*, 59(7): 96–104.

Feuerriegel, S., DiResta, R., Goldstein, J. A., Kumar, S., Lorenz-Spreen, P., Tomz, M., & Pröllochs, N. (2023). Research can help to tackle AI-generated disinformation. *Nature Human Behaviour*, 1-4.

Floridi, L., & Cowls, J. (2019). A unified framework of five principles for AI in society. *Harvard Data Science Review*. https://hdsr.mitpress.mit.edu/pub/l0jsh9d1/release/8

Gigerenzer, G. (2022). *How to stay smart in a smart world: Why human intelligence still beats algorithms*. MIT Press.

Goh, E., Bunning, B., Khoong, E., Gallo, R., Milstein, A., Centola, D., & Chen, J. H. (2023). ChatGPT Influence on Medical Decision-Making, Bias, and Equity: A Randomized Study of Clinicians Evaluating Clinical Vignettes. *Available at https://www.medrxiv.org/content/medrxiv/early/2023/11/27/2023.11.24.23298844.full.pdf*

Goldman, E. (2008). Search engine bias and the demise of search engine utopianism. In: Spink A and Zimmer M (eds) *Web Search: Multidisciplinary Perspectives*. Berlin: Springer, pp.121–133.

Goldstein, J. A., Chao, J., Grossman, S., Stamos, A., & Tomz, M. (2024). How persuasive is AI-generated propaganda? *PNAS Nexus, 3*, pgae034.

Greene, J. D., Sommerville, R. B., Nystrom, L. E., Darley, J. M., & Cohen, J. D. (2001). An fMRI investigation of emotional engagement in moral judgment. *Science*, *293*, 2105-2108.

Grinberg, N., Joseph, K., Friedland, L., Swire-Thompson, B., & Lazer, D. (2019). Fake news on Twitter during the 2016 U.S. presidential election. *Science*, *363*, 374–378.

Haidt, J. (2024). *The anxious generation: How the great rewiring of childhood is causing an epidemic of mental illness*. Penguin Press.

Haridas, G., Sohee, S. K., & Brahmecha, A. (2023). The key policy frameworks governing AI in India. *https://accesspartnership.com/the-key-policy-frameworks-governing-ai-in-india/*

Hoes, E., Altay, S., & Bermeo, J. (2023). Leveraging ChatGPT for efficient fact-checking. Available at https://osf.io/preprints/psyarxiv/qnjkf/

Holt-Lunstad, J. (2021). The Major Health Implications of Social Connection. *Current Directions in Psychological Science, 30*, 251-259.

Holt-Lunstad, J., Smith, T. B., Layton, J. B. (2010). Social Relationships and Mortality Risk: A Meta-analytic Review. *PLOS Medicine, 7*, e1000316.

Ishowo-Oloko, F., Bonnefon, J. F., Soroye, Z., Crandall, J., Rahwan, I., & Rahwan, T. (2019). Behavioural evidence for a transparency–efficiency tradeoff in human–machine cooperation. *Nature Machine Intelligence, 1,* 517-521.

Jobin, A., Ienca, M., & Vayena, E. (2019). The global landscape of AI ethics guidelines. *Nature Machine Intelligence*, *1*, 389-399.

Johnson, M. (2021). A scalable approach to reducing gender bias in Google translate. https://ai.googleblog.com/2020/04/a-scalable-approach-to-reducing-gender.html

Johnson, M., Albizri, A. & Harfouche, A. (2021). Responsible Artificial Intelligence in Healthcare: Predicting and Preventing Insurance Claim Denials for Economic and Social Wellbeing. *Information Systems Frontiers, 25*, 2179-2195.

Johnson, B. S., *et al.* (2023). Using ChatGPT to evaluate cancer myths and misconceptions: artificial intelligence and cancer information. *JNCI Cancer Spectrum, 7*, pkad015.

Jolley, D., & Douglas, K. M. (2017). Prevention is better than cure: Addressing anti-vaccine conspiracy theories. *Journal of Applied Social Psychology*, *47*, 459-469.

Kahane, G., Everett, J. A., Earp, B. D., Caviola, L., Faber, N. S., Crockett, M. J., & Savulescu, J. (2018). Beyond sacrificial harm: A two-dimensional model of utilitarian psychology. *Psychological Review*, *125*, 131-164.

Kanjee, Z., Crowe, B., & Rodman, A. (2023). Accuracy of a Generative Artificial Intelligence Model in a Complex Diagnostic Challenge. *JAMA,* 330, 78–80.

Kasneci, E. et al. (2023). ChatGPT for good? On opportunities and challenges of large language models for education. *Learning and Individual Differences, 103*, 102274.

Köbis, N., Bonnefon, J. F., & Rahwan, I. (2021). Bad machines corrupt good morals. *Nature Human Behaviour*, 5, 679-685.

Korinek, A. (2023). Generative AI for economic research: Use cases and implications for economists. *Journal of Economic Literature*, 61: 1281–1317.

Kranzberg, M. (1985). The information age: Evolution or revolution. *Information technologies and social transformation*, 35-54.

Kreps, S., McCain, R., & Brundage, M. (2022). All the news that's fit to fabricate: AI-generated text as a tool of media misinformation. *Journal of Experimental Political Science, 9*, 104-117.

Kumar, A., & Epley, N. (2021). It's surprisingly nice to hear you: Misunderstanding the impact of communication media can lead to suboptimal choices of how to connect with others. *Journal of Experimental Psychology: General*, *150,* 595-607.

Laufer, R. S., & Wolfe, M. (1977). Privacy as a concept and a social issue: A multidimensional developmental theory. *Journal of Social Issues*, *33*(3), 22-42.

Licklider, J. C. (1960). Man-computer symbiosis. *IRE Transactions on Human Factors in Electronics*, 4-11.

Liu, M., et al. (2023). Future of education in the era of generative artificial intelligence: Consensus among Chinese scholars on applications of ChatGPT in schools. *Future Education Research, 1*, 72–101.

Longoni, C., Bonezzi, A., & Morewedge, C. K. (2019). Resistance to medical artificial intelligence. *Journal of Consumer Research, 46*, 629-650.

Longoni, C., Fradkin, A., Cian, L., & Pennycook, G. (2022). News from generative artificial intelligence is believed less. In *Proceedings of the 2022 ACM Conference on Fairness, Accountability, and Transparency* (pp. 97-106).
44

Madan, R., & Ashok, M. (2023). AI adoption and diffusion in public administration: A systematic literature review and future research agenda. *Government Information Quarterly*, *40*, 101774.

Makovi, K., Sargsyan, A., Li, W., Bonnefon, J. F., & Rahwan, T. (2023). Trust within human-machine collectives depends on the perceived consensus about cooperative norms. *Nature Communications, 14,* 3108.

Mannuru, N.R., Shahriar, S., Teel, Z.A., et al. (2023). Artificial intelligence in developing countries: The impact of generative artificial intelligence (AI) technologies for development. *Information Development, 0(0),* 1–19.

March, C. (2021). Strategic interactions between humans and artificial intelligence: Lessons from experiments with computer players. *Journal of Economic Psychology*, 87, 102426.

McCarthy, B. (June 7, 2023). Ron DeSantis ad uses AI-generated photos of Trump, Fauci. *AFP*. https://factcheck.afp.com/doc.afp.com.33H928Z

McGrew, S. (2024). Teaching Lateral Reading: Interventions to Help People Read like Fact Checkers. *Current Opinion in Psychology, 55*, 101737.

Mei, Q., Xie, Y., Yuan, W., & Jackson, M. O. (2024). A Turing test of whether AI chatbots are behaviorally similar to humans. *Proceedings of the National Academy of Sciences*, *121*, e2313925121.

Merton, R. K. (1948). The Self-Fulfilling Prophecy. *The Antioch Review, 8*, 193-210.

Mill, J. S. (1861/2016). Utilitarianism. In *Seven masterpieces of philosophy* (pp. 329-375). Routledge.

Morgan Lewis (2023). https://www.morganlewis.com/pubs/2023/10/european-trilogue-session-on-eu-ai-act-concludes-with-questions-remaining#:~:text=October%2026%2C%202023,AI%20Act%20in%20May%202023

Natale, S. (2021). *Deceitful Media: Artificial Intelligence and Social Life after the Turing Test*. Oxford University Press.

Natale, S., & Cooke, H. (2021). Browsing with Alexa: Interrogating the impact of voice assistants as web interfaces. *Media, Culture & Society,* 43, 1000-1016.

Nov, O., Singh, N., & Mann, D. M. (2023). Putting ChatGPT's medical advice to the (Turing) Test. *Available at* https://www.medrxiv.org/content/10.1101/2023.01.23.23284735v2

Noy, S., & Zhang, W. (2023). Experimental evidence on the productivity effects of generative artificial intelligence. *Science, 381*, 187–192.

Obermeyer, Z., Powers, B., Vogeli, C., & Mullainathan, S. (2019). Dissecting racial bias in an algorithm used to manage the health of populations. *Science, 366(6464),* 447–453.

Pasquale, F. (2020). *New laws of robotics*. Harvard University Press.

Patel, S. B., & Lam, K. (2023). ChatGPT: the future of discharge summaries? *Lancet Digit Health,* 5, e107–e108.

Peng, S., Kalliamvakou, E., Cihon, P., & Demirer, M. (2023). The impact of AI on developer productivity: Evidence from GitHub Copilot. *Available at https://arxiv.org/abs/2302.06590.*

Pennycook, G., & Rand, D. G. (2022). Accuracy prompts are a replicable and generalizable approach for reducing the spread of misinformation. *Nature Communications*, *13*, 2333.

Penninkilampi, R., Casey, A. N., Singh, M. F., & Brodaty, H. (2018). The association between social engagement, loneliness, and risk of dementia: a systematic review and meta-analysis. *Journal of Alzheimer's Disease*, *66*, 1619-1633.

Peters, U. (2022). Algorithmic political bias in artificial intelligence systems. *Philosophy & Technology, 35*, 25.

Pierson, E., Cutler, D.M., Leskovec, J., Mullainathan, S., & Obermeyer, Z. (2021). An algorithmic approach to reducing unexplained pain disparities in underserved populations. *Nature Medicine, 27,* 136–140.

Pretus, C., Javeed, A. M., Hughes, D., Hackenburg, K., Tsakiris, M., Vilarroya, O., & Van Bavel, J. J. (2024). The Misleading count: an identity-based intervention to counter partisan misinformation sharing. *Philosophical Transactions of the Royal Society B*, *379*(1897), 20230040.

Promberger, M., & Baron, J. (2006). Do patients trust computers?. *Journal of Behavioral Decision Making*, *19*, 455-468.

Public Law Project (2023). Key principles for an alternative AI white paper. https://publiclawproject.org.uk/content/uploads/2023/06/AI-alternative-white-paper-in-template.pdf

Quartz (2017). https://qz.com/911968/bill-gates-the-robot-that-takes-your-job-should-pay-taxes

Quinn, T. P., Senadeera, M., Jacobs, S., Coghlan, S., & Le, V. (2021). Trust and medical AI: the challenges we face and the expertise needed to overcome them. *Journal of the American Medical Informatics Association*, *28*, 890-894.

Rahwan, I. (2018). Society-in-the-loop: programming the algorithmic social contract. *Ethics and Information Technology, 20*, 5-14.

Rahwan, I., Cebrian, M., Obradovich, N., Bongard, J., Bonnefon, J. F., Breazeal, C., ... & Wellman, M. (2019). Machine behaviour. *Nature, 568*, 477-486.

Rai, L., & Chunrao, D. (2016). Influencing factors of success and failure in MOOC and general analysis of learner behavior. *International Journal of Information and Education Technology*, *6*, 262-268.

Rathbone, J. A., Cruwys, T., Jetten, J., & Barlow, F. K. (2020). When stigma is the norm: How weight and social norms influence the healthcare we receive. *Journal of Applied Social Psychology, 53*, 185-201.

Rathje, S., Mirea, D. M., Sucholutsky, I., Marjieh, R., Robertson, C., & Van Bavel, J. J. (2023a). GPT is an effective tool for multilingual psychological text analysis. *Available at https://osf.io/preprints/psyarxiv/sekf5*

Rathje, S., Robertson, C., Brady, W. J., & Van Bavel, J. J. (2023b). People think that social media platforms do (but should not) amplify divisive content. *Perspectives on Psychological Science*, 17456916231190392.

Ren, F., Zhou, Y. (2020). CGMVQA: A new classification and generative model for medical visual question answering. *IEEE Access, 8*, 50626–50636.

Reverberi, C., Rigon, T., Solari, A., Hassan, C., Cherubini, P., & Cherubini, A. (2022). Experimental evidence of effective human–AI collaboration in medical decision-making. *Scientific Reports*, *12*, 14952.

Roberts, H. (2023). The future of AI policy in China. *East Asia Forum.* https://www.eastasiaforum.org/2023/09/27/the-future-of-ai-policy-in-china/

Roozenbeek, J., & Van der Linden, S. (2019). Fake news game confers psychological resistance against online misinformation. *Palgrave Communications*, *5*, 1-10.

Roozenbeek, J., Van Der Linden, S., Goldberg, B., Rathje, S., & Lewandowsky, S. (2022). Psychological inoculation improves resilience against misinformation on social media. *Science Advances*, *8*, eabo6254.

Selwyn, N., Hillman, T., Bergviken Rensfeldt, A., & Perrotta, C. (2023). Digital technologies and the automation of education. *Postdigital Science and Education, 5*, 15-24

Semrush Team. (2023). Maximizing SEO Impact with ChatGPT: A Comprehensive Guide. *Semrush Blog*. https://www.semrush.com/blog/chatgpt-seo/

Septiandri, A.A., Constantinides, M., & Quercia, D. (2023). The impact of AI innovations on U.S. occupations. *Nokia Bell Labs, Cambridge, UK,* work in progress.

Shan, G., & Qiu, L. (2023). Examining the Impact of Generative AI on Users' Voluntary Knowledge Contribution: Evidence from A Natural Experiment on Stack Overflow. *Available at https://papers.ssrn.com/sol3/papers.cfm?abstract_id=4462976*.

Shin, D., & Akhtar, F. (2024). Algorithmic Inoculation Against Misinformation: How to Build Cognitive Immunity Against Misinformation. *Journal of Broadcasting & Electronic Media*, 1-23.

Simchon, A., Edwards, M., & Lewandowsky, S. (2024). The persuasive effects of political microtargeting in the age of generative artificial intelligence. *PNAS nexus*, *3*(2), pgae035.

Spitale, G., Biller-Andorno, N., & Germani, F. (2023). AI model GPT-3 (dis) informs us better than humans. *Science, 9*, eadh1850.

Sternisko, A., Cichocka, A., & Van Bavel, J. J. (2020). The dark side of social movements: Social identity, non-conformity, and the lure of conspiracy theories. *Current Opinion in Psychology*, *35*, 1-6.

Stoyanovich, J., Van Bavel, J. J., & West, T. V. (2020). The imperative of interpretable machines. *Nature Machine Intelligence*, *2*, 197-199.

The Lancet Regional Health – Europe, Embracing generative AI in health care (2023). *The Lancet Regional Health – Europe, 30*.

Topol, E. J. (2019). High-performance medicine: the convergence of human and artificial intelligence. *Nature Medicine, 2*, 44–56.

Twomey, J., Ching, D., Aylett, M. P., Quayle, M., Linehan, C., & Murphy, G. (2023). Do deepfake videos undermine our epistemic trust? A thematic analysis of tweets that discuss deepfakes in the Russian invasion of Ukraine. *Plos One*, *18*, e0291668.

Valtorta, N. K., Kanaan, M., Gilbody, S., Ronzi, S., & Hanratty, B. (2016). Loneliness and social isolation as risk factors for coronary heart disease and stroke: Systematic review and meta-analysis of longitudinal observational studies. *Heart*, *102*, 1009-1016.

Van der Linden, S. (2023). *Foolproof: Why Misinformation Infects our Minds and How to Build Immunity*. New York, NY: WW Norton.

Van Lange, P. A. M., & Columbus, S. (2021). Vitamin S: Why is social contact, even with strangers, so important to well-being?. *Current Directions in Psychological Science*, *30*, 267-273.

von Schenk, A., Klockmann, V., & Köbis, N. (2023). Social Preferences Toward Humans and Machines: A Systematic Experiment on the Role of Machine Payoffs. *Perspectives on Psychological Science*, in press.

Wall Street Journal. (2024). The Deepfake Election Has Arrived. *The Journal Podcast*. https://www.wsj.com/podcasts/the-journal/the-deepfake-election-has-arrived/dc188be9-3039-4ab0-a0ee-792648748480

White House (2023). FACT SHEET: President Biden issues executive order on sage, secure, and trustworthy artificial intelligence. https://www.whitehouse.gov/briefing-room/statements-releases/2023/10/30/fact-sheet-president-biden-issues-executive-order-on-safe-secure-and-trustworthy-artificial-intelligence/#:~:text=October%2030%2C%202023%20FACT%20SHEET%3A,Releases%20Today%2C%20President%20Biden%20is

Wietzke, F.B., & McLeod, C. (2013). Jobs, Wellbeing, and Social Cohesion: Evidence from Value and Perception Surveys. *World Bank Policy Research Working Papers,* 6447.

Wirtz, J., Kunz, W. H., Hartley, N., & Tarbit, J. (2023). Corporate digital responsibility in service firms and their ecosystems. *Journal of Service Research*, *26*, 173-190.

Wu, T. (2016). *The Attention Merchants: The Epic Scramble to Get Inside Our Heads*. PRH Knopf, New York.

Yang, Y., Davis, T., & Hindman, M. (2023). Visual misinformation on Facebook. *Journal of Communication, 73*, 316-328.

Yin, Y., Jia, N., & Wakslak, C. J. (2024). AI can help people feel heard, but an AI label diminishes this impact. *Proceedings of the National Academy of Sciences*, *121*(14), e2319112121.

Zavolokina, L., Sprenkamp, K., Katashinskaya, Z., Jones, D. G., & Schwabe, G. (2024). Think Fast, Think Slow, Think Critical: Designing an Automated Propaganda

Detection Tool. *Proceedings of the CHI Conference on Human Factors in Computer Systems (CHI '24)*.

Zheng, S., Trott, A., Srinivasa, S., Parkes, D. C., & Socher, R. (2022). The AI Economist: Taxation policy design via two-level deep multiagent reinforcement learning. *Science Advances, 8*, eabk2607.

Zuboff, S. (2023). The age of surveillance capitalism. In *Social Theory Re-Wired* (pp. 203-213). Routledge.

# Supplementary Information

| Future research directions | | | |
|---|---|---|---|
| **Research area** | **Specific question** | **Potential design** | **Trade-off** |
| Investigate how AI can be used to make information more accessible, especially for individuals with disabilities. | Can AI-based summarization tools improve information accessibility for individuals with cognitive disabilities by simplifying complex texts? | Evaluate the comprehension of complex news articles by individuals with cognitive disabilities after using AI-based summarization tools, compared with individuals not using these tools. | Summarization could make content more accessible, but oversimplification might omit critical details, challenging the balance between accessibility and content accuracy. |
| | How can AI-generated audio descriptions for visual media impact comprehension and enjoyment for visually impaired users? | Compare the experience of visually impaired users consuming visual media with and without AI-generated audio descriptions, measuring both comprehension and enjoyment levels. | While AI-generated audio descriptions can significantly enhance the accessibility and enjoyment of visual media, overly detailed descriptions could distract or confuse. |
| | Can AI-powered mobility assistants improve navigation in public spaces for individuals with physical disabilities? | Evaluate the impact of AI-powered assistants on the independence and mobility of individuals with physical disabilities in public settings. | AI-powered assistants may increase independence and accessibility to public spaces; but over-reliance on AI assistants might inhibit the |
| | | | development of personal navigation skills. |
| Understand how the largest firms could monopolize the future of AI; find ways for smaller and innovative firms to effectively compete with those largest players. | Does open-source AI decrease the risk of monopolization by large firms? | Compare the growth and success rates of firms that contribute to or use open-source AI versus those relying on proprietary solutions. | Open-source could democratize AI development, but the capacity of small firms to benefit from it may be smaller compared to large firms', affecting the potential of such measures to reduce the competitive gap. |
| | Which regulatory frameworks can effectively counteract the anticompetitive advantages gained through data monopolies? | Evaluate the effectiveness of various regulatory frameworks (e.g., data sharing mandates) in promoting competition. | Regulation could prevent monopolistic behaviors, but overly stringent regulations might stifle growth and innovation, particularly for smaller firms navigating compliance complexities. |
| | How does access to venture capital affect the ability of small AI firms to innovate and compete? | Analyze the correlation between venture capital funding levels for small firms and their subsequent innovation output and market performance. | Venture capital could provide crucial resources for innovation, but the focus on rapid returns might push firms toward short-term gains over sustainable innovation. |
| Explore regulatory measures to prevent misuse or inappropriate access to data by AI systems. | Can a standardized transparency protocol prevent data misuse and privacy breaches? | Develop a standardized transparency protocol and test its impact on the frequency of data | Increased transparency may decrease data misuse and privacy breaches, but the difficulty of |

|  |  | misuse and privacy breaches. Measure also users' trust in the system and data sharing. | transparency protocols may actually decrease public trust and data sharing. |
|---|---|---|---|
|  | What is the effect of data anonymization techniques on data misuse and personalization of information? | Test various data anonymization techniques and measure their impacts on data misuse, privacy breaches, and personalization of AI-filtered information. | Data anonymization techniques can reduce data misuse and privacy breaches, but may also reduce the personalization and usefulness of AI-filtered information. |
|  | How do opt-in versus opt-out data consent models affect user willingness to share data with AI systems for personalized content filtering? | Experiment with opt-in and opt-out consent models, measuring changes in user data sharing behavior and the effectiveness of personalized content filtering. | Opt-in models might increase user trust and consent quality, but result in less data being shared, possibly reducing the personalization effectiveness of AI systems. |
| Investigate strategies to identify and limit the spread of misinformation generated by AI. | Can dialoguing with gen AI reduce beliefs in misinformation? | Compare misinformation beliefs between participants engaging with an AI-system designed to counter misinformation and those interacting with a neutral AI. | AI's capabilities for personalization, linguistic fluency, and logical reasoning can diminish misinformation beliefs, but the risk of hallucinations could generate new misinformation. |
|  | Can machine learning models trained on AI-generated text effectively distinguish between human and AI- | Train machine learning models on datasets containing human and AI-generated texts, testing their ability to | Unique linguistic fingerprints left by AI might enable effective differentiation, but AI's ability to mimic human writing styles |
| | generated misinformation? | accurately classify unseen texts. | challenges detection efforts with potential negative consequences in case of misclassification of misinformation. |
| | Can verification systems on social media and messaging platforms reduce the spread of bot-generated misinformation? | Introduce a user verification system to distinguish between human and bot accounts on messaging platforms, analyzing the impact on misinformation spread in one-to-one communications. | Verification could reduce bot-driven misinformation, but the very existence of verification systems may decrease trust in the platform. |
| Explore ways to design AI-systems that support cooperative and ethical behavior in human-machine interactions. | Do humans become more unethical when they can delegate decisions to AI compared to other humans? | Subjects make morally ambiguous decisions, with the option to delegate these decisions either to AI or to another human, observing changes in their ethical standards. | The emotional detachment of AI may reduce moral responsibility, but the potential for AI to document and expose decisions might increase personal accountability. |
| | What features of AI systems and users' understanding of them promote human-AI cooperation? | Compare the extent of human cooperation with AI-systems in social dilemma situations, experimentally manipulating relevant interface design features and system descriptions. | Well-designed AI systems may improve cooperation, but cooperation may not reach the level observed in human-human interaction. |
| | How do different cultural perceptions of ethics influence the design of AI systems for global | Design AI systems with ethical guidelines informed by diverse cultural norms. Test these | Cultural customization could increase the global applicability and acceptance of AI |

|  | human-machine interaction? | systems in various cultural contexts, assessing acceptance and effectiveness in promoting ethical interactions. | systems, but reconciling conflicting ethical norms poses significant challenges for universal AI design. |
|---|---|---|---|
| Examine how AI-enhanced search engines can be designed to preserve user autonomy and plurality of information. | Does making the data sources of AI-enhanced search engines transparent influence user engagement with alternative information sources? | Disclose the data sources for AI algorithms in search engines and assess whether users seek out additional, alternative sources of information as a result. | Transparency about data sources might encourage users to seek diverse information, but could also lead to information overload. |
|  | How do interactive AI assistants that guide users in refining search queries influence the range of information explored? | Deploy interactive AI assistants that help users refine their queries and track the breadth of information they explore compared to users without assistance. | Interactive guidance could expand the scope of information users consider, but reliance on suggestions might limit independent exploration. |
|  | What role does user education on search engine algorithms play in preserving information plurality? | Implement an educational program about how AI search algorithms work and measure its impact on users' ability to seek out diverse information sources. | Educating users could enhance their ability to find diverse information, but overemphasizing algorithmic literacy may burden users without addressing underlying biases in the algorithms. |

| Consider how the proliferation of AI-generated content could lower the quality of online information and ensure that human users can continue to contribute new knowledge. | How does the ratio of AI-generated to human-generated content affect content diversity on online forums? | Manipulate the ratio of AI-generated to human-generated content in an online forum environment. Survey users on perceived content diversity and quality. | Higher ratios of AI content could enhance content availability, but may also homogenize the content or lead to mediocre or over-creation of banal, "average" content. |
|---|---|---|---|
| | How does the ratio of AI-generated to human-generated content influence human users' motivation to contribute original content? | Manipulate the ratio of AI-generated to human-generated content and measure the extent to which humans contribute original content over time. | A higher proportion of AI content could provide users with more information, potentially increasing creativity, but also potentially saturating the platform and demotivating human contributors. |
| | Can AI content curation algorithms be trained to prioritize and highlight innovative human-generated content over derivative AI-generated pieces? | Develop an AI curation algorithm that differentiates between innovative human-generated content and derivative AI content. Measure its impact on the visibility of human contributions. | Prioritizing human creativity enhances content diversity and innovation, but algorithmic bias towards human-generated content could ignore valuable AI contributions or insights. |
| Investigate the role of Corporate Digital Responsibility and its implementation challenges | What are the main barriers to implementing effective digital responsibility strategies in organizations? | Development of organizational strategies to implement a general Corporate Digital Responsibility framework. | Many organizations might not be aware of the role and importance of corporate digital responsibility strategies, which might cause over-regulation. |

| | How does the cost-benefit analysis of digital responsibility practices influence organizational commitment? | Develop predictive models to forecast the long-term benefits and costs associated with digital responsibility practices. | Stakeholders may be wary of decisions made on the basis of analysis if the process is not transparent or if the rationale behind predictions and recommendations is not clearly understandable. |
|---|---|---|---|
| | In what ways can regulation support or hinder the adoption of good corporate digital responsibility practices? | Engage in comparative policy analysis and legal research to understand the landscape of existing regulations affecting corporate digital responsibility practices across different jurisdictions. | While regulatory analysis can provide a detailed understanding of the legal framework, it might not fully capture the operational and strategic challenges faced by organizations in complying with these regulations. |

*Table S1. Summary of the main research directions on the impact of generative AI on information. For each relatively broad research area, we propose three specific research questions, along with potential experimental designs. We also identify an underlying theoretical trade-off that complicates the derivation of a priori hypotheses. This list is not exhaustive but rather provides examples of the kinds of questions future research should aim to address.*

| \multicolumn{4}{c}{**Future research directions**} | | | |
|---|---|---|---|

| **Research area** | **Specific question** | **Potential design** | **Trade-off** |
|---|---|---|---|
| Investigate how AI can be designed and implemented to augment human skills and increase productivity, rather than to simply replace workers and forego the long-run productivity upsides of maximizing workers' contributions to production. | How do organizations balance automating knowledge tasks versus hiring more knowledge workers for efficiency and innovation? | Analyze organizational decisions, assessing the ratio of investment in automation technologies to the recruitment of knowledge workers. | Automation offers cost efficiency and 24/7 productivity, but the unique insights brought by human knowledge workers present a compelling case for augmentation over full automation. |
| | What is the role of AI in enhancing the creative process for designers and artists without diminishing originality? | Artists and designers use AI tools in a controlled study, with assessments of creativity, efficiency, and originality, compared to a control group not using AI. | Balancing AI's input with human creativity involves complex interactions that could affect the originality of outputs unpredictably. Additionally, bias in training data (e.g., Western oriented) might diminish creativity. |
| | Does the use of generative AI free up human resources for more strategic tasks within organizations? | Organizations implement generative AI tools in a pilot program and track the allocation of human resources to strategic tasks versus before AI implementation. | The assumption that AI frees up human resources could be challenged by the need for increased oversight and quality control. |
| Examine how AI can facilitate more access to economic opportunities, particularly through | What role does AI play in enabling marginalized groups, including countries in the Global South, to | Analyze employment trends and outcomes in marginalized groups before and after introducing AI- | AI could provide anonymized job matching, enhancing access, but restrictions on technology use |
| reducing language-related barriers and promoting remote work technologies that can democratize access to the digital economy. | access remote work opportunities in the digital economy? | enabled remote work platforms. | could negate these benefits. |
| | Does generative AI reduce language barriers in cross-cultural collaboration, especially with regard to low-resource language? | Organizations introduce AI-driven language translation tools in some teams but not others, measuring collaboration effectiveness and project outcomes. | AI improves communication but might limit personal language growth and cultural insight. For low-resource languages, scant training data can increase translation errors. |
| | Can AI-powered job matching platforms more effectively connect remote workers with global employment opportunities compared to traditional platforms? | Compare employment success rates and job satisfaction between remote workers using AI-powered job matching platforms and those using traditional platforms. | AI's ability to match skills with opportunities may improve job market efficiency, but biases in AI algorithms and digital divide could perpetuate inequalities. |
| Conduct long-term studies to monitor the evolving impact of AI on the workforce, capturing both the immediate and delayed effects on work across educational and occupational strata. | How does continuous use of AI tools impact job satisfaction and employee burnout? | Track job satisfaction and burnout rates among employees using AI tools versus those using traditional methods over multiple years. | AI tools might boost innovation and creativity, but could also lead to higher expectations and workload, decreasing satisfaction and increasing burnout. |
| | Can AI-driven educational tools reduce the skill gap between workers from different educational backgrounds? | Implement AI-based educational tools and measure skill levels before and after intervention across workers with varying backgrounds. | AI personalized learning could upskill workers, but differences in digital literacy and access may widen the skill gap. |
| | What are the long-term effects of AI on work-life balance | Longitudinal survey assessing work-life balance pre- and post- | AI might offer more flexible working conditions, but |
| | across different occupational sectors? | AI adoption in varying sectors. | increased monitoring and job demands could negatively impact work-life balance. |
| Explore how AI can be utilized in educational and training programs to encourage basic competency with generative AI tools and better-equip workers in vulnerable job sectors in anticipation of labor market changes. | How does intergenerational collaboration within teams, specifically focused to the exchange of AI tool knowledge, impact team performance, inclusivity perceptions, and psychological well-being? | Create diverse teams to assess the impact of intergenerational AI knowledge-sharing on performance, inclusivity perception, and well-being. | Intergenerational knowledge sharing might improve team performance, inclusivity perceptions, and well-being, but it could also introduce challenges such as resistance to change from senior members or frustration among juniors, with negative effects on the outcome variables. |
| | Can AI-based personalized learning platforms increase the effectiveness of re-skilling programs for workers facing automation? | Workers facing job displacement are enrolled in re-skilling programs using AI-based personalized learning, compared to traditional re-skilling approaches. | Personalization may accelerate skill acquisition, yet differences in learning styles and technology familiarity among workers can produce a re-skilling gap. |
| | How effective is generative AI in bridging the digital literacy gap for older workers participating in workforce re-entry programs? | Older workers in re-entry programs use generative AI tools aimed at improving digital literacy, comparing outcomes to those in programs without AI support. | AI tools could provide personalized, pace-adjustable learning, but pre-existing technology apprehensions among older workers may hinder engagement and effectiveness. |
| Research labor laws, taxation policies, and social support systems that could support | How do worker retraining and job replacement programs impact the economic | Incorporate AI-skills into pilot worker retraining programs in regions with high rates | AI-centric worker retraining could enable workers to be more agile in finding |

| workers displaced or disadvantaged by AI. | stability of workers displaced by AI? | of AI-driven displacement, comparing against control groups in similar regions with alternative, non-AI worker retraining. | re-employment in the event of job displacement, but this may not be successful for all workers, particularly those later in their careers or with fewer pre-existing technical skills. |
|---|---|---|---|
| | How do changes in labor laws to protect workers influence their job security and income stability in AI-intensive industries? | Analyze workers' job security and income before and after implementing new labor protections in AI-heavy sectors, comparing with sectors where laws remain unchanged. | Enhanced labor protections could improve job security and income, but increased operational costs for employers might reduce the availability of work. |
| | What are the effects of AI on wage inequality within sectors, and can progressive taxation of AI-generated profits mitigate these effects? | Analyze wage distributions within sectors before and after the introduction of progressive taxation on AI profits, using control sectors without such taxation for comparison. | Progressive taxation might redistribute wealth and reduce inequality, but could also discourage AI investment and innovation, affecting sector growth and wage levels. |

*Table S2. Summary of the main research directions on the impact of generative AI in workplace environments, along with specific example questions and corresponding experimental design and theoretical trade-offs.*

| **Future research directions** | | | |
| --- | --- | --- | --- |
| **Research area** | **Specific question** | **Potential design** | **Trade-off** |
| Examine how generative AI can be effectively used for personalized learning. | How does the use of generative AI for personalized feedback in essay writing impact students' writing proficiency over a semester? | Compare students in classes receiving personalized AI feedback with those receiving standard teacher feedback. Assess improvements in writing proficiency. | Personalized AI feedback could offer more tailored and immediate improvements, but may lack the psychological understanding and motivational impact of human feedback. |
| | How do generative AI-driven simulations impact students' understanding of abstract scientific concepts? | Science classes incorporate generative AI-driven simulations to teach abstract concepts, comparing students' conceptual understanding and engagement to classes using traditional teaching methods. | Simulations could visually and interactively convey complex concepts. However, over-reliance on simulations might limit abstract understanding. |
| | What is the long-term effect of using generative AI for homework assistance on students' independent learning skills? | Students using generative AI tools for homework assistance are tracked over an academic year, comparing their development of independent learning skills to peers who do not use AI assistance. | AI tools could provide personalized help and boost learning efficiency, but might also reduce students' initiative to tackle challenges and limit the development of independent learning skills. |
| Investigate how curricula can be redesigned to include generative AI as a tool | How does integrating generative AI in science curricula affect students' | Implement generative AI in a subset of science classes, comparing students' | Generative AI may enhance concept understanding through personalized learning, |
| for enhancing learning while also teaching students to critically engage with and understand this technology. | understanding of complex concepts? | performance on concept understanding to those in classes without AI tools. | but the potential for over-reliance on AI could hinder independent critical thinking skills. |
| | Can generative AI tools assist in teaching problem-solving skills more effectively than conventional tools? | Compare the effectiveness of generative AI tools versus conventional tools in teaching problem-solving. | The adaptive learning algorithms of AI could offer customized problem-solving practice, but the potential for hallucinating reasonable but wrong mathematical solutions could hinder learning. |
| | How does the use of generative AI for creating interactive history lessons affect students' historical empathy? | Measure changes in historical empathy in students engaging with interactive, AI-generated history lessons versus traditional textbook-based learning. | Interactive content could deepen engagement and understanding, but overemphasizing technology might detract from human-based discussion and reduce historical empathy. |
| Study effective training methods for teachers to integrate AI tools into their teaching practices and identify the additional support required to manage these technologies in the classroom. | Can peer mentoring programs increase teachers' confidence in using generative AI tools? | Implement a peer mentoring program where AI-experienced teachers mentor those less familiar, and assess changes in confidence and usage rates of AI tools in teaching. | Peer mentoring could provide useful support and encouragement, but mismatches in teaching styles might limit the program's effectiveness and even reduce confidence. |
| | How does collaborative training in AI tool integration affect teachers' | Split teachers into groups receiving either collaborative or individual training on | Collaborative training could enhance teamwork and collective problem- |

|  | classroom implementation success compared to individual training methods? | AI tool integration, followed by evaluations of their success rate of AI tool implementation in their classrooms. | solving skills, but individual differences in learning pace and style might lead to uneven skill acquisition and implementation success. |
|---|---|---|---|
|  | Can just-in-time training methods enhance teachers' ability to integrate AI tools into their curriculum? | Implement a just-in-time training program for a group of teachers, and compare their integration success to teachers who received traditional training sessions. | Just-in-time training might offer immediately applicable knowledge, but the lack of comprehensive training could leave teachers unprepared for broader integration challenges. |
| Explore strategies to ensure that use of generative AI leads to a diversity of educational experiences and outcomes | How does the educational use of generative AI affect students' creativity? | Study three groups of students: one using generative AI with educator support for creativity, one using AI independently, and the other not using AI. Assess the creativity of their outputs. | Generative AI can expand students' creativity with proper guidance but might reduce originality and insight if used without educational support. |
|  | How does the educational use of generative AI support access to a plurality of knowledge, viewpoints, and perspectives? | Test the effect of pedagogically advanced uses of generative AI on alternate viewpoints, divergent and critical thinking. | Generative AI can offer a broad spectrum of ideas and stimulate critical thinking with the right pedagogical approach, it risks reinforcing limited perspectives inherent in its training data if not carefully managed. |
|  | How does the educational use of generative AI support the inclusion of | Classrooms adopt AI tools and practices designed for inclusivity, measuring | AI tools could enhance accessibility and learning for socially disadvantaged |

| | learners who are otherwise marginalized and disadvantaged? | the academic and social integration outcomes of otherwise disadvantaged students compared to classrooms without such tools. | students and those with disabilities, but implementation challenges and the need for teacher training might even decrease integration outcomes. |
|---|---|---|---|
| Evaluate the long-term impacts of generative AI on student learning, teacher workloads, and educational outcomes. | Does the use of generative AI in lesson planning reduce teachers' preparation time? | Track lesson preparation time for teachers using generative AI tools versus traditional planning methods over a semester. | AI tools could streamline the lesson planning process, but the initial learning curve and adjustments to AI suggestion might offset time savings. |
| | How does long-term use of generative AI affect students' critical thinking skills? | Compare critical thinking skill development over a school year between students taught with generative AI and those with a traditional curriculum. | Generative AI could provide more engaging and diverse content, potentially enhancing critical thinking, but might also lead to overreliance. |
| | How does AI-based personalized learning impact long-term student motivation and academic persistence? | Track student motivation and persistence in academic programs using generative AI personalization versus traditional programs, over several years. | Personalized learning could boost motivation and persistence by aligning with students' interests and needs, but potential overreliance might diminish intrinsic motivation. |

*Table S3. Summary of the main research directions on the impact of generative AI on education, along with specific research questions, with corresponding experimental design and theoretical trade-offs.*

| \multicolumn{4}{c}{**Future research directions**} |

| **Research area** | **Specific question** | **Potential design** | **Trade-off** |
|---|---|---|---|
| Research how AI can assist healthcare professionals in diagnosis, treatment planning, and patient monitoring. | Does AI-assisted medical decision making improve patient outcomes including decision accuracy and bias reduction? | Healthcare providers are given the option to consult or not consult medical advice from generative AI chatbots, with decision accuracy and decision bias measured across medical contexts and healthcare provider characteristics. | Clinicians could heed the advice of AI, improving their medical decision making. However, mistrust for AI tools, low AI literacy, or other sources of suboptimal interactions with AI could render AI-assisted decision making ineffective or detrimental for patient outcomes. |
| | How does AI-assisted diagnosis compare to traditional diagnostic methods in terms of diagnosis accuracy? | Professionals diagnose using traditional methods in one group and AI assistance in another, comparing diagnostic accuracy. | AI could enhance diagnostic accuracy, due to its ability to analyze patterns in data, but over-reliance might overcome the clinician's experiential intuition, potentially affecting diagnosis accuracy. |
| | How does the integration of AI in patient monitoring impact the management of post-operative recovery? | Post-operative patients are monitored using AI-based systems in one group and traditional monitoring methods in another, comparing recovery rates, | AI monitoring could provide real-time data and early warning signs of complications, enhancing recovery management. However, reliance |

| | | complications, and readmission rates. | on technology might reduce direct patient-clinician interactions, potentially impacting care quality. |
|---|---|---|---|
| Investigate the use of AI to reduce the administrative burden on healthcare providers through efficient electronic health records (EHR) management. | Does the integration of AI into EHR systems reduce the time healthcare providers spend on documentation compared to traditional EHR systems? | Healthcare providers are divided into two groups, one using AI-integrated EHR systems and the other using traditional EHR systems, with the time spent on documentation measured over a specified period. | AI could streamline documentation processes, reducing time spent on paperwork. However, learning curves associated with new AI systems might initially increase the time required for documentation. |
| | Can AI-powered speech-to-text solutions enhance the accuracy of clinical documentation in EHRs compared to manual typing by healthcare providers? | Providers use AI-powered speech-to-text for clinical documentation in one group, while another group relies on manual typing. The accuracy and completeness of records are compared. | Speech-to-text could improve efficiency and accuracy by reducing typing errors. However, speech recognition errors could introduce inaccuracies, affecting record quality. |
| | Can AI in EHRs reduce healthcare costs associated with administrative tasks? | Analyze financial records to compare administrative costs before and after AI implementation in EHR systems. Include both direct costs and indirect costs. | AI could lower long-term costs by increasing efficiency, but the upfront costs of AI technology and necessary training could be substantial. |
| Study how AI can contribute to the development of | Can AI-based predictive models accurately identify | Develop AI models using patient health data to predict | Predictive models could allow for earlier and more |
| personalized medicine, adapting treatments to individual patient needs and reducing healthcare disparities. | patients at high risk for diabetes (or any other disease) and guide early intervention strategies? | diabetes risk and apply intervention strategies based on model predictions. Compare the incidence of diabetes in this group with a control group receiving standard care. | targeted interventions. However, inaccuracies in predictions could lead to unnecessary interventions or miss at-risk individuals. |
| | How effective is AI in optimizing vaccine distribution strategies to maximize coverage and equity in diverse populations? | Use AI to design vaccine distribution strategies in diverse populations, comparing coverage and equity to strategies developed through traditional methods. | AI could optimize distribution logistics, enhancing coverage and equity. However, algorithmic biases might inadvertently exacerbating disparities. |
| | Can AI-enhanced remote patient monitoring systems reduce hospital readmission rates? | Compare readmission rates of patients monitored using AI-enhanced systems with those of patients monitored using standard remote systems. | Enhanced monitoring might detect early signs of deterioration, reducing readmissions. Yet, overreliance might delay seeking in-person care when needed, potentially risking patient health. |
| Investigate strategies to increase public trust and understanding of AI in healthcare. | Does providing patients with detailed explanations of AI diagnostic processes improve their trust in AI-driven healthcare? | Patients receiving AI-driven diagnostics are split into two groups, one receiving detailed explanations and the other receiving standard information. Trust levels are measured post-interaction. | Detailed explanations could demystify AI processes, increasing trust. However, overly technical or complex information might overwhelm patients, potentially reducing trust. |

| | What role does personalized communication from AI healthcare systems play in patient satisfaction and trust? | Implement AI healthcare systems that use personalized communication with patients, comparing patient satisfaction and trust to systems using generic communication. | Personalized communication might foster a sense of care and understanding. Yet, if perceived as insincere, it could diminish trust and satisfaction. |
|---|---|---|---|
| | How does media campaign depicting AI's role in healthcare influence public perception and trust? | Analyze public perception and trust in AI before and after targeted media campaigns depicting AI's role in healthcare, comparing with a control group not exposed to the campaigns. | Positive media representation could improve public perception and trust in AI, but sensationalism or negative portrayals in some media could exacerbate fears and distrust. |
| Research how AI can improve healthcare accessibility in underserved regions and populations, in both rural and urban areas. | Can AI-driven mobile health applications effectively increase healthcare access in rural communities? | Implement an AI-driven mobile health application in rural communities. Compare health outcomes, access to care, and self-management behaviors with a control group not using the application. | Mobile health apps could significantly improve access and self-management, but poor internet connectivity and low digital literacy in rural areas might limit effectiveness and even widen disparities. |
| | Can AI algorithms improve the accuracy of disease outbreak predictions in underserved areas, leading to better preparedness and response? | Develop AI algorithms to predict disease outbreaks in underserved areas, comparing the accuracy of predictions and subsequent response efforts with historical outbreaks managed | Improved prediction accuracy could significantly enhance outbreak response. Yet, reliance on incomplete or biased data sets might result in inaccurate predictions, potentially |

|  |  | without AI predictions. | misguiding response efforts. |
|---|---|---|---|
|  | How effective are AI-based systems in detecting early signs of mental health issues in underserved communities? | Use AI-based systems to monitor and detect early signs of mental health issues in underserved communities, compared with traditional detection methods. | AI might offer early and accurate detection, but misinterpretations or privacy concerns could hinder the trust and effectiveness of these systems. |
| Investigate the potential of AI to facilitate social connections, particularly for individuals with difficulties in forming relationships, while also studying the potential risks of over-reliance on AI for social interaction. | Can AI-based chatbots effectively reduce feelings of loneliness compared to traditional social programs? | Participants are divided into two groups, with one interacting with AI-based chatbots and the other engaging in traditional social programs. Measures of loneliness and social skills are assessed over time and their influence on health. | AI agents could offer constant companionship, potentially reducing loneliness. However, overreliance on AI could inhibit motivation for human interaction leading to increased isolation. Further, the lack of genuine human interaction might not fulfill deep social needs, affecting overall health. |
|  | How can AI be used to increase the quality of digital social interaction and communication? | Compare the human-to-human digital exchanges (emails, texts, DM, etc) when AI pro-social prompts are offered to those without prompts. | AI could prompt people to be more respectful, kind and courteous. The con is that people may view such messages as less authentic. |

|  | How does reliance on AI for social interaction impact the development of social skills in adolescents? | Track social skill development in adolescents who heavily rely on AI for social interaction versus those who primarily engage in human interactions, assessing communication skills, empathy, and relationship quality. | AI interaction might offer continuous social practice, but the absence of complex human feedback from multiple social companions could flatten social skills. |
|---|---|---|---|

*Table S4. Summary of the main research directions on the impact of generative AI on healthcare, along with specific example questions.*